\documentclass[useAMS,usegraphicx,usenatbib]{mn2e}
\usepackage{graphicx,aas_macros,amsmath,verbatim,rotating,color,natbib}
\usepackage{fixltx2e}[1999/12/01]

\newcommand{\kpc}{ \ensuremath{ h^{-1} {\rm kpc}} }

\newcommand{\nopt}{\ensuremath{ N_{\rm opt}}}
\newcommand{\lx}{\ensuremath{ L_{\rm X}}}
\newcommand{\sz}{\ensuremath{Y_{\rm SZ}}}
\newcommand{\mlens}{\ensuremath{M_{\rm lens}}}
\newcommand{\planck}{\small Planck}
\newcommand{\MCXC}{\small MCXC}
\newcommand{\REXCESS}{\small REXCESS}

\begin{document}
\setlength{\topmargin}{-1.cm}
\title[]
{Scaling relations for galaxy clusters in the Millennium-XXL simulation}

\author[Angulo et al.]{
\parbox[h]{\textwidth} {R. E. Angulo$^{1}\thanks{rangulo@mpa-garching.mpg.de}$, V.~Springel$^{2,3}$,
  S.~D.~M.~White$^{1}$, A.~Jenkins$^{4}$, C.~M.~Baugh$^{4}$,
  C.~S.~Frenk$^{4}$} \vspace*{6pt} 
\\ 	
\\ $^1$ Max-Planck-Institute for Astrophysics, Karl-Schwarzschild-Str. 1, 85740 Garching, Germany.
\\ $^2$ Heidelberg Institute for Theoretical Studies, Schloss-Wolfsbrunnenweg
35, 69118, Heidelberg, Germany,
\\ $^3$ Zentrum f\"{u}r Astronomie der
Universit\"{a}t Heidelberg, ARI, M\"onchhofstr. 12-14, 69120 Heidelberg,
Germany.
\\ $^4$ Institute for Computational Cosmology, Dep.~of Physics,
Univ.~of Durham, South Road, Durham DH1 3LE, UK } \maketitle

\date{\today}
\pagerange{\pageref{firstpage}--\pageref{lastpage}} \pubyear{2011}
\label{firstpage}

\begin{abstract} 
  We present a very large high-resolution cosmological N-body
  simulation, the {\em Millennium-XXL} or MXXL, which uses 303 billion
  particles to represent the formation of dark matter structures
  throughout a $4.1$~Gpc box in a $\Lambda$CDM cosmology.  We create
  sky maps and identify large samples of galaxy clusters using
  surrogates for four different observables: richness estimated from
  galaxy surveys, X-ray luminosity, integrated Sunyaev-Zeldovich
  signal, and lensing mass. The unprecedented combination of volume
  and resolution allows us to explore in detail how these observables
  scale with each other and with cluster mass.  The scatter correlates
  between different mass-observable relations because of common
  sensitivities to the internal structure, orientation and environment
  of clusters, as well as to line-of-sight superposition of
  uncorrelated structure. We show that this can account for the
  apparent discrepancies uncovered recently between the mean thermal
  SZ signals measured for optically and X-ray selected clusters by
  stacking data from the Planck satellite.  Related systematics can
  also affect inferences from extreme clusters detected at high
  redshift. Our results illustrate that cosmological conclusions from
  galaxy cluster surveys depend critically on proper modelling, not
  only of the relevant physics, but also of the full distribution of
  the observables and of the selection biases induced by cluster
  identification procedures.
\end{abstract}
\begin{keywords}
cosmology:theory - large-scale structure of Universe.
\end{keywords}

\section{Introduction}

Confrontation of observational data with theoretical models, and in
particular with numerical simulations, has been a key factor enabling
the rapid recent progress of cosmological research. Without it, we
would have not arrived at the current structure formation paradigm,
which is now being subjected to ever more detailed scrutiny.  Further
development of this fruitful approach requires current and future
observations to be matched with equally precise theoretical
models. The order of magnitude advances made by new surveys hence
require new simulations with comparable improvements in statistical
power and accuracy.

An inevitable consequence of the increasing accuracy of observational
data and the growing sophistication of numerical simulations is that
comparing them becomes a non-trivial task in its own right.  It has
long been appreciated that the distribution of properties in a sample
of observed objects is shaped not only by the relevant physics but
also by the observational methods used to detect and characterise
them.  The resulting measurement biases have often been neglected in
the past, but this is no longer possible in the era of `precision
cosmology' where the systematic errors in observational results are
typically comparable to or larger than their statistical errors.
Detailed modelling of a given observational programme is not optional
in this situation, but rather is an unavoidable step in the proper
interpretation and exploitation of the data.

In this paper, we present a major new effort in this direction, aiming
to address two aspects of the physics of galaxy clusters that have
recently attracted a lot of interest.  The first concerns the
relations between optical richness, lensing mass, X-ray luminosity and
thermal Sunyaev-Zeldovich (tSZ) signal. It is critically important
to calibrate how these observables scale with 'true' mass if cluster
counts are to be used to place robust constraints on cosmological
parameters.  The {\planck} Collaboration has recently reported
puzzling inconsistencies in the scaling relations measured for
different samples, suggesting an unexpected dichotomy in the gas
properties of galaxy clusters \citep{PlanckOptical2011}.  The other
aspect concerns the inferred masses of extreme galaxy clusters.  This
is interesting because discoveries of massive clusters at high
redshift have repeatedly been suggested to be in tension with the
standard $\Lambda$CDM model, possibly providing evidence for 
non-Gaussian initial density perturbations
(\citealp{Mullis2005,Hoyle2011,Foley2011,Brodwin2010,Baldi2011,Menanteau2011,
Hoyle2012}, but see \citealp{Hotchkiss2011}).

To study these questions, we use a new state-of-the-art simulation of
the evolution of the dark matter structure that provides the arena for
the formation and evolution of galaxies. This {\em Millennium-XXL} is
the largest high-resolution cosmological N-body simulation to date,
extending and complementing the previous Millennium and Millennium-II
simulations \citep{Springel2005b,Boylan-Kolchin2009}. It follows the
dark matter distribution throughout a volume equivalent to that of the
whole sky up to redshift $z=0.7$, or equivalently, of an octant up to
redshift $z=1.4$. Its time and mass resolution are high enough to
allow detailed modelling of the formation of the galaxy populations
targeted by future large surveys, as well as of the internal
structure of extremely rare and massive clusters. It is also well
suited for studying a number of other probes of the cosmic expansion
and structural growth histories, for example, baryonic acoustic
oscillations (BAOs), redshift space distortions, cluster number
counts, weak gravitational lensing, and the integrated Sachs-Wolfe
effect.  The volume of Millennium-XXL is very much larger than can be
followed by direct hydrodynamical computations, but its resolution is
sufficient for galaxy formation to be followed in detail within each
halo by applying semi-analytic models to its merger tree.

Using mock observations of galaxy clusters in the Millennium-XXL, we
show in this study that current interpretations of cluster surveys are
significantly affected by systematic biases.  In particular, we show
that the apparent inconsistencies highlighted by the {\planck}
Collaboration in the mean SZ and X-ray signals measured for optically
and X-ray selected cluster samples can be understood as resulting from
substantial and correlated scatter in the various observables among
clusters of given 'true' mass.  Currently there appears to be no
compelling evidence for unknown processes affecting the gas properties
of clusters or for a bimodality in cluster scaling properties. We also
comment on the implications of our results for constraining cosmology
using extreme clusters at high redshift.

Our paper is structured as follows. Section~2 is devoted to presenting
the Millennium-XXL and our techniques for modelling the observable
properties of galaxy clusters. In particular, Section~2.1 provides
technical and numerical details of the simulation, while Section~2.2
describes our surrogates for X-ray luminosity, gravitational lensing
mass, optical richness and thermal SZ flux. In Section 3, we then
explore the impact and implications of various selection biases. We
explain how we identify clusters in Section~3.1, and in Section~3.2 we
analyse the bulk of the cluster population, with a focus on extreme
objects. We discuss the implication of our findings for the conundrum
reported by the {\planck} Collaboration in Section~4. Finally, we
present our conclusions in Section~5.

\section{Numerical methods}

In this section we describe our dark matter simulation and the way we
use it to construct surrogates for four observational properties of
galaxy clusters; their optical richness, their weak gravitational lensing
signal, their X-ray luminosity and their thermal Sunyaev-Zeldovich (tSZ) amplitude.

\subsection{The MXXL N-body simulation}

\begin{figure*}
\centering
\resizebox{17cm}{!}{\includegraphics{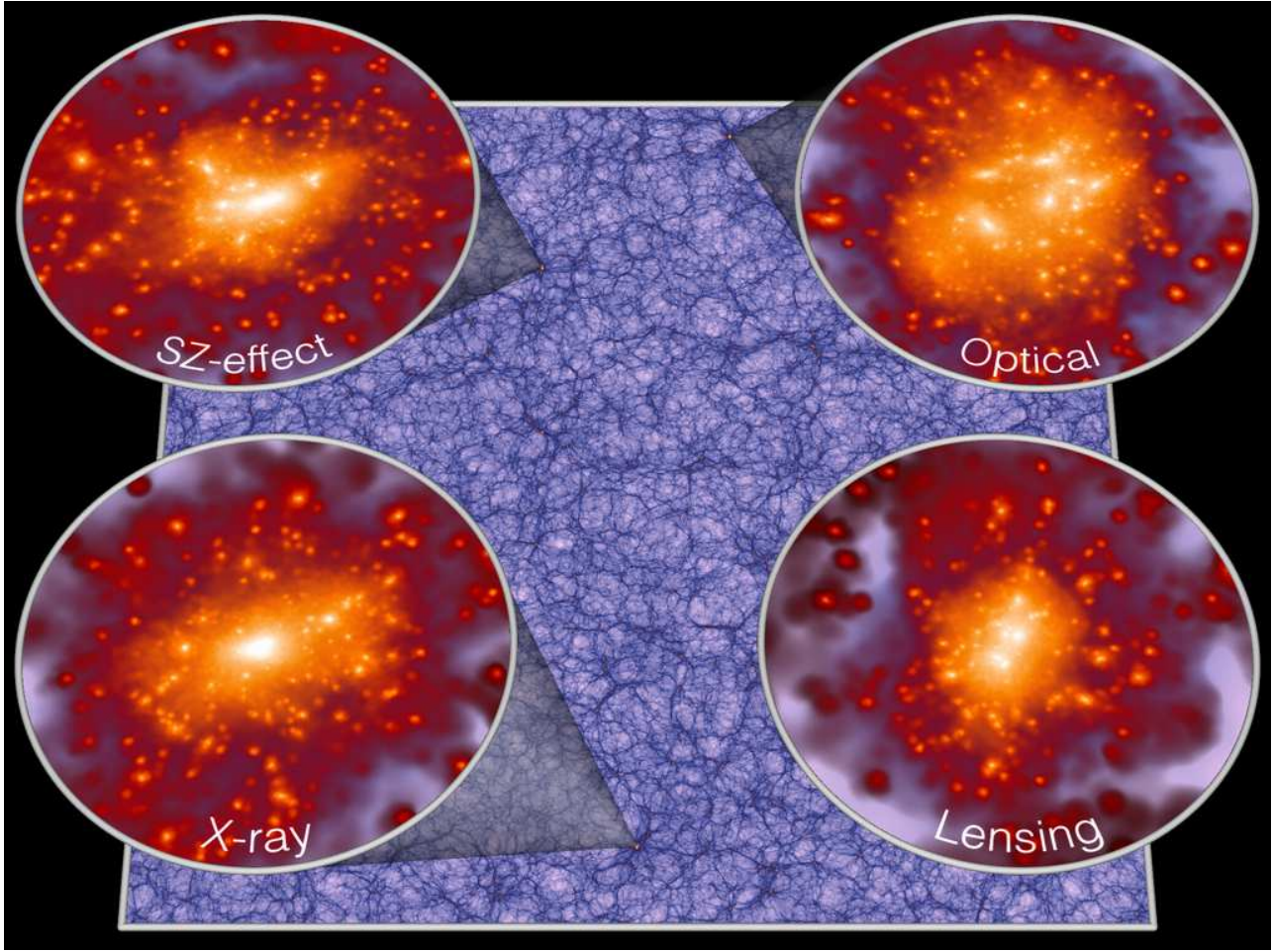}}
\caption{The projected density of dark matter in the MXXL simulation
  at $z=0.25$. The insets correspond to circles of radius
  $5.5\,{\rm Mpc}$ centred on the most extreme clusters identified
  according to our surrogates for X-ray luminosity, optical richness,
  lensing signal and integrated thermal tSZ strength (see
  section~\ref{sec:surrogates}).  The underlying image is a
  projection of the dark matter density in a slab of thickness
  $27\,{\rm Mpc}$, and width $2050\,{\rm Mpc}$. It is oriented so that
  it contains three of the selected clusters, as indicated in the
  figure (the lensing example is from a different slice). The whole simulation box is actually twice as wide, spanning
  $4110\,{\rm Mpc}$.  All four cluster images and the large-scale
  slice use the same colour scale, which varies in shade from light
  blue in the least dense regions to orange and white in the densest
  regions. }
\label{mxxl}
\end{figure*}

The `Millennium-XXL Simulation' (MXXL) follows the nonlinear growth of
dark matter structure within a cubic region of $4.11\,{\rm Gpc}$
($3\,h^{-1}{\rm Gpc}$) on a side. The dark matter distribution is
represented by $6720^3 = 303,464,448,000$ particles, substantially
exceeding the number used in all previous simulations of this type
\citep{Springel2005a,Kim2009,Teyssier2009,Prada2011} apart from the recent
`Horizon Run 3' of \citet{Kim2011}, which has 20\% more particles but
40 times poorer mass resolution. We note that the simulated volume of the
MXXL is equivalent to that of the whole observable Universe up to redshift
$z=0.72$. It is more than $200$ times that of our `Millennium
Simulation' \citep[MS,][]{Springel2005a}, almost 30 times that of the
recently completed MultiDark simulation \citep{Prada2011} but still
only 2\% that of the Horizon Run 3. The MXXL is also about $7$ times
larger than the expected volume of the Baryon Oscillation
Spectroscopic Survey (BOSS) \citep{Schlegel2007} and about twice that 
of the planed JPAS\footnote{http://www.j-pas.org}. Its particle mass
is $m_{p} = 8.456 \times 10^{9}\, {\rm M}_{\odot}$,
approximately $7$ times that of the MS but more than $300$
times smaller than that of the `Hubble Volume Simulation'
\citep{Evrard2002}, completed a decade ago with a comparable volume to
MXXL. The mass resolution of MXXL is sufficient to identify the dark matter
haloes hosting central galaxies 
with stellar mass exceeding $\sim 1.5\times10^{10}\,{\rm M}_\odot$
\citep{deLucia2006}, and also
to predict robustly the internal properties of the haloes
corresponding to very massive clusters, which are represented by more
than $100,000$ dark matter particles.  The Plummer-equivalent
softening length of the gravitational force is $\epsilon=13.7\,{\rm kpc}$, 
which translates into a dynamic range of $300,000$ per
dimension, or formally to more than $2\times10^{16}$ resolution elements
within the full simulation volume. This large dynamic range can be
appreciated in Fig.~\ref{mxxl}, where we show the large-scale
density field together with the internal structure of a few selected
massive clusters.

The MXXL adopts a $\Lambda$CDM cosmology with the same cosmological
parameters and output times as the previous two Millennium simulations
\citep{Springel2005a,Boylan-Kolchin2009}. This facilitates the
joint use of all these simulations in building models for the galaxy population.  Specifically,
the total matter density, in units of the critical density, is
$\Omega_{\rm m} = \Omega_{\rm dm} + \Omega_{\rm b} = 0.25$, where
$\Omega_{\rm b} = 0.045$ refers to baryons (although these are not
explicitly treated in the simulation); a cosmological constant,
$\Omega_\Lambda = 0.75$, gives a flat space geometry; the {\it rms}
linear density fluctuation in $10.96\,{\rm Mpc}$ spheres, extrapolated
to the present epoch, is $\sigma_8 = 0.9$; and the present-day Hubble
constant is $H_0 = 73\,{\rm km\,s^{-1} Mpc^{-1}}$. Although this set
of parameters is discrepant at about the $3\sigma$ level with the
latest constraints from CMB and LSS observations \citep{Komatsu2011},
the scaling technique proposed by \cite{Angulo2010} \citep[see
 also][]{Ruiz2011} allows the Millennium simulations to provide
theoretical models for the formation, evolution and clustering of
galaxies over the full range of cosmologies allowed by current
observational constraints. The parameter offset with respect to the
best current observational estimates lies mainly in the high value for
$\sigma_8$, but this is an advantage for reliable scaling of
the simulation results to other cosmologies, as this requires interpolation on
the stored MXXL/MS/MS-II data which is only possible for target
cosmologies with lower $\sigma_8$ than used in the MS.

\subsubsection{Initial conditions}

The initial unperturbed particle load for the simulation was built by
periodically replicating a $280^3$ particle cubic glass file
twenty-four times in each coordinate direction. The glass file was
created for the MXXL using the method of \cite{White1996}
\citep[see also][]{Baugh1995}.  The initial conditions were then
produced by computing displacements and velocities for each of the
particles at starting redshift $z_{\rm start}=63$, using an
upgraded version of the code originally developed for the Aquarius
Project \citep{Springel2008}.  Further improvements include
communication and memory optimisations, as well as the use of
second-order Lagrangian perturbation theory \citep[2LPT,][]{Scoccimarro1998},
rather than the Zel'dovich approximation for computing the position
and velocity perturbations of each particle. The latter modification
is particularly important for the present study, since the abundance of high
mass haloes is sensitive to initial transients, which
are much smaller and decay more quickly when 2LPT is used
\citep{Crocce2006}.

Another important change is the introduction of a new approach to
generate Gaussian initial fluctuations.  Rather than setting the
phases of the modes in $k$-space (as done, for example, in the MS), we
first generated a real-space white noise field. The Fourier transform
of this field was then used to set the amplitudes of all of the modes
needed to make the initial conditions
\citep{Salmon1996,Bertschinger2001,Hahn2011}.  For the MXXL, the
real-space white noise field was created on a $9216^3$ grid.  Only
modes within a spherical $k$-space volume of radius $6720/2 = 3360$
times the fundamental frequency (i.e. below the particle Nyquist
frequency) were used to generate the displacement and velocity fields
(all other modes were given zero amplitude).

The use of a white noise field in real space, while not necessary for
the MXXL initial conditions themselves, will make it much
easier to resimulate arbitrary MXXL regions of interest at higher
resolution, for example, the extreme objects illustrated in the present
paper.  This is because in our new approach it is unnecessary to
reproduce the entire white noise field at the original resolution in
order to capture the phases of large-scale modes. A consequence is the
ability to create consistent sets of initial conditions for resimulations (including
`resimulations of resimulations' at yet higher resolution) for arbitrarily 
defined subregions over a huge dynamic range. The real-space white noise field is
generated in a special top-down hierarchical fashion, based on an
oct-tree, making it easy to generate coarse representations of the
MXXL field at low computational cost. The MXXL white noise
field itself occupies just a small subvolume of a single realisation of a huge
white noise field created in a hierarchical way. This realisation is specified
everywhere to a resolution below the likely free streaming scale of cold dark 
matter. This means that resimulations of parts of the MXXL volume can be 
created at any desired resolution as the phases are fully specified everywhere 
in advance. A full description of this method will be given in Jenkins (2012, 
in preparation).

\subsubsection{The simulation code}

Evolving the distribution of the dark matter particles in the MXXL
under their mutual gravitational influence was a formidable
computational problem.  Storing the positions and velocities of the
particles in single precision already requires about $7\,{\rm TB}$ of
memory. As each particle exerts a force on every other particle, a
CPU- and memory-efficient approximate calculation of the forces is of 
paramount importance. It is also necessary to
develop new strategies to deal with the huge data volume produced by
the simulation. Using the same analysis approach as for the
Millennium Simulation would have resulted in more than $700\,{\rm TB}$ of data,
adding a severe data analysis problem and significant disk space costs
to the computational challenge.

In order to alleviate these problems, we developed a special
``lean'' version of the Tree-PM code {\small GADGET-3}, which improves
the scalability and memory efficiency of the code considerably,
outperforming the highly optimised version of {\small GADGET-2}
\citep{Springel2005b} used for the MS.  {\small GADGET-3}
computes gravitational forces with a TreePM method by combining a
particle-mesh (PM) scheme with a hierarchical tree-method, and it uses
spatially and temporally adaptive time-stepping, so that short
time-steps are used only when particles enter localised dense
regions where dynamical times are short.  A significant
improvement in the new code is a domain decomposition
that produces almost ideal scaling on massively parallel computers.
Finally, the MXXL version of {\small GADGET-3} uses aggressive
strategies to minimise memory consumption without compromising
integration accuracy and computational speed.  
To be specific: (i) We have taken advantage of the unused bits in the 64-bit
particle IDs to store various quantities during the calculation, e.g. 4/8-bit
floats containing the number of interactions and the  acceleration in the
previous timestep (these allow us to improve the work-load balance), the
group/subhalo membership, and the time-step bin. (ii) Each MPI task
contains multiple disjoint sequences of the Peano-Hilbert curve describing the
computational domain, resulting in an almost perfect load decomposition. (iii)
We avoid storing the geometric center for each node in the tree
structure used to
compute gravitational forces, reducing the memory requirement at the cost
of a slightly less effective tree opening criterion. We also note that we have
searched for the combination of force and time-integration parameters that
minimises the total execution time for a given desired accurcy in the simulation results.

The code also carries
out a significant part of the required post-processing on-the-fly as
an integral part of the simulation. This includes group-finding via
the Friends-of-Friends \citep{Davis1985} (FOF) algorithm, application
of the {\small SUBFIND} algorithm \citep{Springel2001b} to find
gravitationally bound subhaloes within these groups, and calculation
of basic properties of these (sub)haloes, like maximum circular
velocities, cumulative density profiles, halo shapes and orientations, velocity
dispersions, etc. These extended halo and subhalo catalogues are then
stored at the same output times as for the other Millennium
simulations, allowing the construction of detailed (sub)halo merger
trees. Full particle data are, however, stored only at a handful of
redshifts, very significantly reducing the stored data volume.  Only
72 bytes per particle are needed by the simulation code during 
normal dynamical evolution. When the in-lined group and substructure
finders are enabled as well (which is optional), the peak memory
consumption per particle increases by a further 26 bytes.

\subsubsection{Computational cost and code performance}

The MXXL simulation was carried out in the late summer of 2010 on the
JuRoPa machine at the J\"ulich Supercomputing Centre (JSC) in Germany. A partition
of 1536 compute nodes was used, each equipped with two quad-core Intel
X5570 processors and 24 GB of RAM.  We ran our code in a hybrid
MPI/shared memory setup on $12,288$ cores, placing one MPI task per
processor socket (3072 in total), and employing all four cores of each
socket via threads. This setup turned out to be advantageous compared
with a pure MPI-parallelisation based on 12888 MPI tasks, because it
reduces the amount of intra-node MPI communication, and minimises the
RAM required for MPI communication buffers. Also, this makes it easier
for our code to reach close-to-optimum work- and load-balance during the
calculation.  

The final production run carried out approximately 87 trillion force
calculations to reach $z=0$, and used about $28.5\,{\rm TB}$ of RAM,
nearly the whole available physical memory of JuRoPa. The run-time was
$9.3$ days (wall-clock), equivalent to $2.86$ million CPU hours (or
$326$ years) in serial. Of this time, 15\% were required for running
our on-the-fly postprocessing software, notably the group finding, the
substructure finding, and the power spectrum calculation, and another
14\% were needed for I/O operations.  The total long-term storage
space required for all MXXL data products is about $100\,{\rm TB}$,
down by a factor of about 8 per particle relative to the approach used
for the MS and MS-II simulations.

\subsection{Basic validation results}

\begin{figure}
\centering
\resizebox{8.5cm}{!}{\includegraphics{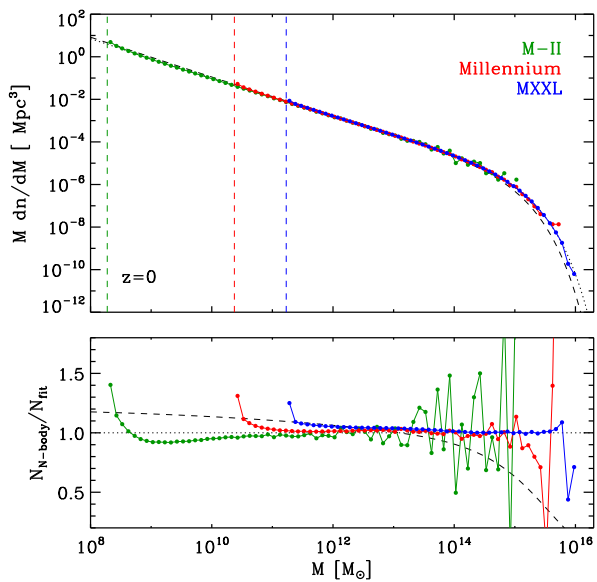}}
\caption{ The differential Friends-of-Friends (FOF) halo mass function
  (top panel) of the MXXL (blue), MS (red) and MS-II (green).
  The MXXL provides vastly superior sampling of the
  massive end, where the abundance of objects drops exponentially as a
  function of mass. Combined, the three simulations cover about 8
  decades in halo mass.  The vertical lines mark the halo resolution
  limits (20 particles) of the three simulations. For comparison, we also 
  display a fit to
  the mass function of all self-bound subhaloes in the three Millennium
  simulations (dashed). The bottom panel
  gives the ratio of the three mass functions to an analytic fitting
  formula given in the text.  We see that the simulations agree accurately
  with each other for intermediate masses, but also that different
  methods for identifying structures disagree significantly in the
  expected number density of objects of given mass, especially at the
  high-mass end.
}
\label{FigMF}
\end{figure}

\begin{figure}
  \centering
  \vspace*{-0.5cm}\resizebox{8.5cm}{!}{\includegraphics{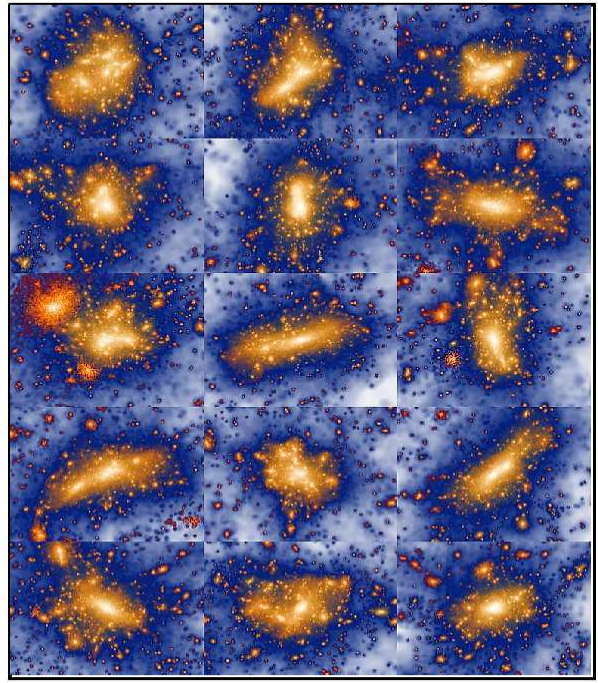}}
  \vspace*{-0.9cm}\caption{Projected dark matter density for the 15
    most massive MXXL haloes (according to $M_{200}$ at $z=0.25$. Each
    image corresponds to a region of dimensions $6\times3.7\,h^{-1}$Mpc wide and 
    $20\,h^{-1}$Mpc deep. Note the large variation in shape and internal structure
    among these clusters. In particular, the most massive cluster, shown in the
    top left corner, has no clear centre but rather displays several distinct
    density peaks of similar amplitude.}
\label{plot:clusters}
\end{figure}

\begin{figure}
\centering
\resizebox{8.5cm}{!}{\includegraphics{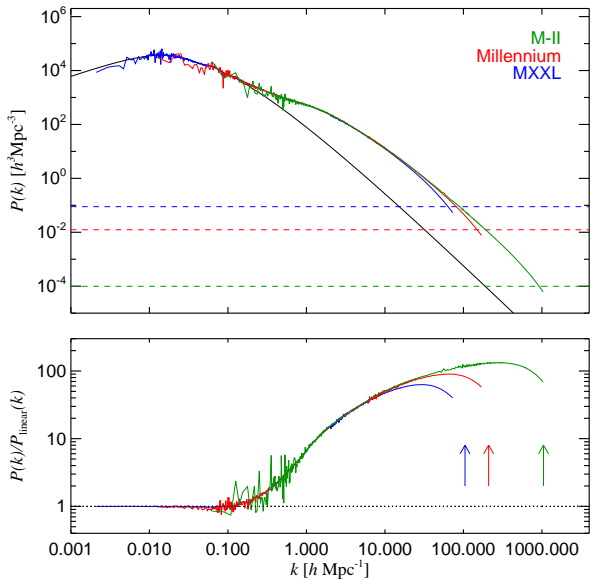}}
\caption{ Matter power spectra measured directly in the MXXL, MS, and
  MS-II (top panel). The black line shows the power
  spectrum used to generate the initial conditions, linearly evolved
  to $z=0$. The dashed lines show the Poisson power level of each
  simulation, which becomes significant only at the smallest
  scales. The Poisson power has been subtracted from the measured
  power spectra in this figure.  In the bottom panel, we show the
  ratio of the measured power spectra to the actual realisation of the
  linear theory used to generate the initial conditions of each
  simulation.  This procedure reduces sampling noise due to the finite
  number of modes at small wavenumber.  The arrows
  mark the gravitational resolution limits ($2\pi/$softening length)
  of the three simulations. }
\label{FigPower}
\end{figure}

At redshift $z=0$, the MXXL contains more than 700 million haloes with
at least $20$ particles. These account for $44\%$ of all the mass in
the simulation.  Among these objects, $23$ million have a value of
$M_{200}$\footnote{We define the conventional virial mass of a halo
  $M_{200}$ as the mass within a sphere centred on the potential
  minimum which has mean density 200 times the critical value.} larger
than that of the Milky-Way's halo ($M_{200} = 2\times10^{12}\,{\rm
  M}_{\odot}$) and $464$ have a value in excess of that of the Coma
galaxy cluster ($M_{200} = 2\times10^{15}\,{\rm M}_{\odot}$).  In
Fig.~\ref{FigMF}, we show the differential halo mass function (FoF
masses for a linking length $b=0.2$) at the present epoch, which is a
robust way of describing the abundance of nonlinear
objects as a function of mass \citep{Davis1985}.  The most massive
halo at $z=0$ has $M_{\rm FoF}=8.98 \times 10^{15}\,{\rm M}_\odot$. 
Such extreme objects are so rare that they can only be
found in volumes as large as that of the MXXL. We compare the MXXL 
results with similar measurements from the MS and MS-II simulations.
For masses where the three simulations have good statistics and are 
away from their resolution limits, the agreement is at the few percent level. The
results from all three simulations are well described by

\begin{equation}
M \frac{{\rm d}n}{{\rm d}M} = \rho_0 \frac{ {\rm d} \ln \sigma^{-1}}{{\rm d}M } f(\sigma(M)),
\end{equation}
where $\rho_0$ is the mean mass density of the universe, $\sigma(M)$ 
is the variance of the linear density field within a top-hat filter containing 
mass $M$, and $f(\sigma)$ is the fitting function
\begin{equation}
f(\sigma(M))= 0.201\times\left[\frac{2.08}{\sigma(M)} + 1\right]^{1.7}\, \exp\left[\frac{-1.172}{\sigma^2(M)}\right].
\end{equation}

\noindent The residuals from this analytic halo mass function, displayed in the
bottom panel of Fig.~\ref{FigMF}, show that it describes the
numerical results accurately (to better than 5\% over most of the mass range) 
over eight orders of magnitude in halo mass,
extending the accuracy of previous models to larger and to smaller
scales \citep[e.g.][]{Jenkins2001,Warren2006,Tinker2008}. In
Fig.~\ref{FigMF} the dashed lines show a fit of this same analytic
form to the mass function of all self-bound subhaloes (as identified by
{\small SUBFIND}) in the MXXL, MS and MS-II simulations. These curves
correspond to the fit

\begin{equation}
f(\sigma(M)) = 0.265\times\left[\frac{1.675}{\sigma(M)} + 1\right]^{1.9}\, \exp\left[\frac{-1.4}{\sigma^2(M)}\right].
\end{equation}
The difference between the two fits illustrates how the mass function
of objects depends on the way they are defined. This is especially
important at the high-mass end. For example, the expected abundance of
haloes with $M\sim10^{15}\,{\rm M}_{\odot}$ changes by a factor of
$\sim 2$ when FoF haloes and self-bound subhaloes are compared.

The difficulty in unambiguously defining haloes and their associated
mass is in part a consequence of the fact that large haloes do not
form a homogeneous population. In fact, they display considerable
variety in structure and environment. We illustrate this in
Fig.~\ref{plot:clusters} by showing the fifteen most massive
clusters in the MXXL at $z=0.25$, selected according to
$M_{200}$. Among this group there is considerable diversity in shape,
concentration and the amount of substructure, despite all the objects
having very similar virial mass. This already suggests that careful
modelling of mass estimators will be needed to compare numerical
simulations with observed massive clusters at high redshift. Small
changes in the estimated mass of an object can dramatically change the
predicted probability of its existence within any given cosmological
model.

The diversity of massive clusters may also have important consequences
for other observational studies. Matched filters are often applied to data
in order to maximise the signal-to-noise ratio of, for instance, weak
lensing or tSZ detections
\citep[e.g.][]{Schneider1996,Melin2006,Rozo2011}.  Such filters use a
model for the spatial distribution of the signal as prior information
(e.g.~in the form of density or pressure profiles) but in many cases
(and in particular for the most massive objects) the structure of
individual clusters will not conform to these assumptions. For instance
the top left halo in Fig.~\ref{plot:clusters}, which is the most
massive cluster in the MXXL at $z=0.25$, does not have a clear
centre.  In such cases, the signal may be seriously misestimated by a
matched filter, potentially biasing cosmological inferences from the
measurement.

In Fig.~\ref{FigPower}, we show power spectra of the mass density
field at the present epoch. The results are a combination of two
measurements.  Large-scale modes were computed using a global $9216^3$
mesh, whereas the mean amplitude of smaller modes was calculated by
folding the density field $64$ times along each direction and
projecting it onto a new $9216^3$ mesh \citep{Jenkins1998}. This method effectively
reaches the same spatial resolution as a $589,824^3$ mesh.  For
comparison, we also show results for the MS and MS-II simulations.
Clearly, only the MXXL simulation probes scales significantly beyond
the turnover in the power spectrum. The MXXL is also the only 
one among the three runs that provides good sampling of the baryonic 
acoustic oscillations (BAO). We note that at low redshift these features 
already show clear signs of being affected by nonlinear evolution 
\citep[e.g.][]{Angulo08}, making the MXXL particularly valuable for
studying systematic effects in large-scale galaxy surveys aiming at precise
measurements of the BAO features. Throughout the
nonlinear regime, the power spectra of the three Millennium
simulations show excellent agreement up to the scales where the
spatial resolution limits of each run kick in, manifested as a
reduction in power relative to higher resolution simulations. 

\subsection{Surrogate observables}\label{sec:surrogates}

Using the DM distribution and halo catalogues described in the
previous section we have created surrogate observables that mimic the
four main techniques used observationally to discover and characterise
large clusters: optical galaxy counts, gravitational lensing, X-ray
emission, and the Sunyaev-Zel'dovich signal imprinted on the microwave
background radiation.

Rather than attempting to follow the baryonic physics directly in the
simulation, we have constructed simple proxies for these observables,
based directly on the dark matter distribution. This necessarily
schematic approach avoids the uncertainties of any specific
implementation of baryonic processes such as star and black hole
formation and the associated feedback, while allowing us to take
advantage of the characteristics of the MXXL, namely its combination of
very large volume and relatively high mass resolution. Our approach
can easily be updated as a better understanding of the relation between
the dark matter structure of galaxy clusters and any particular
observable is achieved. Our main goal in this paper is not to produce
accurate {\it a priori} predictions for the observables, but to look
for surrogates that correctly rank the expected signal strengths and
represent the scatter and the correlations between observables in a
realistic way.  We can then study the diversity of clusters and
quantify the extent to which different methods select different 
cluster populations.

We focus our analysis on redshift $z=0.25$ because the most massive
halo in the observable Universe should be roughly at that redshift
\citep{Holz2010}. This redshift also corresponds to the median
redshift of galaxies in the photometric catalogue of the Sloan
Digital Sky Survey \citep[SDSS][]{York2000}, which provides one of the
largest samples of optically detected clusters -- the MaxBCG catalogue
of \citet{Koester2007b}, which has been widely used to compute scaling
relations for optically selected clusters. Our results do not change
qualitatively if we pick another redshift between $z=0$ and $1$.
The largest simulated haloes should resemble the most massive
observable objects because the MXXL volume is comparable to or exceeds
that accessible to real surveys. In $1000$ all-sky light-cones up to
$z = 0.6$ (built by placing observers randomly on 'Milky Way' haloes),
we find the most massive halo is typically between $z=0.1$ and $z=0.3$, 
and has a virial mass $M_{200} \sim 4 \times10^{15}\,{\rm
  M}_\odot$, roughly 75\% that of the most massive MXXL halo at
$z=0.25$, consistent with previous analytic estimates
\citep{Holz2010}.

Finally, we note that we normalise our surrogates to match observed
scaling relations between observables and halo mass (which we discuss
in Section~4).  This allows a direct comparison of population
properties to observational data, side-stepping issues of possible
offsets due to incorrect cosmological parameters, to the schematic
nature of our surrogates, or to observational details such as filter
shapes.  We now outline how we construct 2D maps from which we can
identify clusters and measure our various surrogate observables.

\subsubsection{Optical maps}

The first observational approach we consider is the detection of rich
clusters in optical surveys, which relies on finding large groups of
galaxies in a narrow redshift range and at similar projected positions
on the sky.  In order to mimic this, we start by constructing a
three-dimensional galaxy catalogue using a halo occupation
distribution (HOD) model \citep{Kauffmann1997,Benson2000,Peacock2000}
to populate each MXXL halo with galaxies. We assume that every halo
with $M_{200}$ above $1.4\times10^{12}\,{\rm M}_\odot$ hosts one
central galaxy and a number of satellites drawn from a Poisson
distribution with mean equal to the halo mass in units of
$5.7\times10^{12}\,{\rm M}_\odot$ raised to the $0.9$ power.  This HOD
is similar to that derived for red galaxies in the SDSS \citep{Zehavi2011}, 
but it is tuned to reproduce the observed mass-richness
relation for galaxy clusters as measured by \cite{Johnston2007}. The central 
galaxy is placed at the minimum
of the gravitational potential of the dominant {\small SUBFIND}
substructure in the halo, whereas the satellite galaxies are
identified with randomly chosen dark matter particles of the FoF
group. The latter ensures that the effects of halo ellipticity,
alignment and substructure, which are important for reproducing the
small-scale correlations of galaxies \citep{Zu2008,Daalen2012}  are
included in our modelling. The resulting catalogue at $z=0.25$
contains more than $150$ million galaxies (50\% of which are
satellites).

We note that the basic assumption of this HOD modelling, namely that
the galaxy content of a halo is statistically determined exclusively
by its mass, is not expected to hold in detail. Models that follow
galaxy formation explicitly predict a dependence of the HOD on other
properties such as the halo formation time
\citep[e.g.][]{Gao2005,Zhu2006,Croton2007}. Nevertheless, these effects are
weak so we do not expect them to affect our conclusions. Similarly, moderately
different HODs (e.g. models tuned to reproduce the \cite{Rozo2009}
mass-richness relation) change the average number of galaxies in our
clusters but make little difference to the correlation of optical properties
with other aspects of the halo.

\subsubsection{Lensing maps}

The second identification approach we consider is weak gravitational
lensing. Although direct mass measurements using this effect are only
possible for the most massive individual clusters, lensing can be used to
estimate precise mean masses by stacking a large number of clusters
selected according to specific criteria \citep[e.g.][]{Mandelbaum2006,Sheldon2009}.  
It is easy to see that orientation effects and both correlated and uncorrelated 
large-scale structure 
can play an important role in defining the lensing signal of an individual
cluster. As a result, large N-body simulations where the full
line-of-sight density distribution is properly modelled are needed to
calculate accurately the distribution of expected signals.

Our modelling of lensing maps uses the distant observer
approximation. We neglect the evolution of clustering along the
line-of-sight and assume that all the mass along a line-of-sight from
$z=0$ to $z=1.37$ (the side length of the MXXL) contributes equally to
the convergence field.  Under these assumptions we create ``weak
lensing mass maps'' by projecting the simulated mass density of the
$z=0.25$ snapshot along one axis. MXXL particles are mapped onto a
$32,768^2$ mesh using a nearest grid point (NGP) mass assignment
scheme, yielding an effective transverse spatial resolution of $\sim
92\,\kpc$.

A more realistic approach would vary the weight assigned to mass at
different redshifts assuming a specific redshift distribution for the
background source galaxies. The highest weight would go to material
which is ``halfway'' to the sources. Our 4.1~Gpc projection length
will clearly tend to overestimate projection effects from distant
matter. However, it turns out that structures far in front or far
behind the clusters produce only a small fraction of the projection
effects; most come from their immediate surroundings and from the
directional dependence of the projection of their internal
structure. Our simple model can be regarded as treating these aspects
quite accurately and as giving an overestimate of the (subdominant)
effects of distant projections.

\subsubsection{X-ray maps}

Another important route to detecting and characterising massive
clusters is through X-ray emission from their hot intracluster gas.
The local X-ray emissivity is proportional to the square of the gas
density and (approximately) to the square root of its temperature.
This makes X-rays a particularly sensitive tracer of the inner regions
of clusters, where densities can be thousands of times higher than in
the outskirts. Simulations with high spatial and mass resolution are
needed to probe these inner regions adequately.

We estimate a density local to each particle by using
kernel-interpolation over its 32 nearest neighbours, a common method
in SPH calculations, while we take the local temperature to be
proportional to the velocity dispersion of the subhalo in which the
particle is located. Particles outside subhaloes are taken to have zero
temperature. With these quantities in hand, we compute a $32,768^2$
pixel X-ray map by summing up the density times the square root of the
temperature for all the particles along a given line of sight.
We note that our X-ray surrogate corresponds to the
total bolometric luminosity of a cluster rather than to the luminosity in
a particular observational band. However, this has little impact on our
results because of the low redshift of our sample and the fact that we scale
our surrogates to match observations.

This surrogate clearly neglects dynamical effects on cluster
luminosity during violent cluster mergers. Recent hydrodynamical
simulations \citep{Rasia2011} suggest that luminosity enhancements
during such events can be substantial, particularly for high Mach
number ($>2.5$) and for equal progenitor masses. On the other hand,
observational data suggest that disturbed, apparently merging clusters
tend to have lower than average X-ray luminosities for their mass,
whereas symmetric equilibrium clusters often have cold cores and thus
higher than average luminosities \citep{Arnaud2010}. Even the sign of
merger effects thus seems unclear.

\subsubsection{Thermal SZ maps}

The final cluster property we consider is their thermal
Sunyaev-Zel'dovich (tSZ) signal \citep{SZ1972,SZ1980}. This effect
causes a characteristic distortion of the spectral shape of the cosmic
microwave background (CMB) as a result of inverse Compton scattering
of CMB photons off the electrons in the hot intracluster plasma. The integrated
magnitude of the effect is proportional to the total thermal energy
content of the hot electrons in the cluster, or, equivalently, to the
gas mass times the mean gas temperature. Along any given line-of-sight
the effect is proportional to the line integral of the gas pressure.

In our analysis, we again assume the gas to be distributed like the
dark matter and to be isothermal within each quasi-equilibrium subhalo,
associating a temperature to each simulation particle proportional to
the velocity dispersion of its host subhalo. We then create a
$32,768^2$ pixel tSZ map by projecting the thermal energies of all
MXXL particles along one of the box axes.

\section{Results}

In this section we use our simulated halo catalogues and mock
observational maps to examine various systematic effects that can
have an impact on the scatter and mean amplitude of cluster scaling relations.
We cumulatively include effects due to sample selection, to spurious
cluster identification resulting from projection effects, to
misidentification of cluster centres, and to contamination by
structures along the line-of-sight, examining in each case the impact
on relations between mean tSZ and optical richness, and between weak
lensing mass and optical richness.

\subsection{Cluster catalogues}\label{catalogues}

We first specify how we identify optical clusters in our galaxy
map. We have implemented a group finder similar to that employed to
build the MaxBCG cluster catalogue from SDSS galaxies
\citep{Koester2007b}. We start by measuring $N_1$, the number of
galaxies within a cylinder of radius $1\,h^{-1}{\rm Mpc}$ and depth
$120\,h^{-1}{\rm Mpc}$ centred on every central galaxy in our
catalogues (the depth mimics a redshift uncertainty of $\Delta z \sim
0.02$). Then we discard those galaxies whose cylinder overlaps with
that of a central galaxy of a more massive cluster. This is equivalent
to assuming that the luminosity of the central Brightest Cluster
Galaxy (BCG) increases monotonically with mass, and then discarding as
potential group centres those BCGs that are close to a brighter
galaxy. After this cleaning procedure we use the galaxy counts around the 
remaining central galaxies to define a new ``observed'' cylinder radius 
$\overline{R}(N_{\rm opt})$, equal to the mean virial radius of clusters of 
the same $N_1$ richness. Then, we repeat the counting and
cleaning processes until we reach convergence. We keep all clusters
down to a count of one galaxy (i.e. just the central BCG). We refer to
this count as the optical richness $N_{\rm opt}$ of the cluster.

A serious systematic in optical cluster catalogues may be caused by
misidentification of the BCG, and hence of the center of the
corresponding dark matter halo \citep{Rozo2011}. This effect is
referred to as `miscentering'. In fact, 20\% to 40\% of the MaxBCG
groups (depending on the cluster richness) suffer from this effect
according to \citet{Johnston2007}. In order to mimic this in our
analysis, we have carried out our cluster identification procedure
{\it after} randomly displacing $30\%$ of our candidate cluster
centres according to a 2D Gaussian with a mean shift of
$0.4\,h^{-1}\rm{Mpc}$. This distribution of offsets is based on
\cite{Johnston2007}, who applied the MaxBCG algorithm to mock
catalogues built from the Hubble simulation \citep{Evrard2002}. The
functional form and parameters we use to describe the effect are also
consistent with the distribution of projected distances between the 
position of the dominant subhalo and that of the second most massive
subhalo within haloes of MXXL simulation
\citep[see also Fig.~2 of ][]{Hilbert2010}. We caution that
the miscentering fraction and the displacement parameters are
uncertain and are sensitive to details of the cluster-finding
algorithm. In \citet{Planck2011} the large X-ray cluster compilation
of \citet{Piffaretti2011} was matched to the maxBCG catalogue, finding
a median offset between X-ray centre and BCG position of about
100~kpc for the 189 clusters in common; $\sim 15\%$ are offset by more
than 400~kpc (J-B Melin, private communication). This agrees
reasonably with the result of \citet{Johnston2007}. We will see below
that few of our results are sensitive to miscentering because it
generally causes cluster observables to be perturbed parallel to the
scaling relations which link them. Since we retain a flag which notes
which of our clusters have had their centres displaced, we can
construct a cluster sample based on ``true'' centres simply by
ignoring these objects.

\begin{figure*}
\centering
\resizebox{8.8cm}{!}{\includegraphics{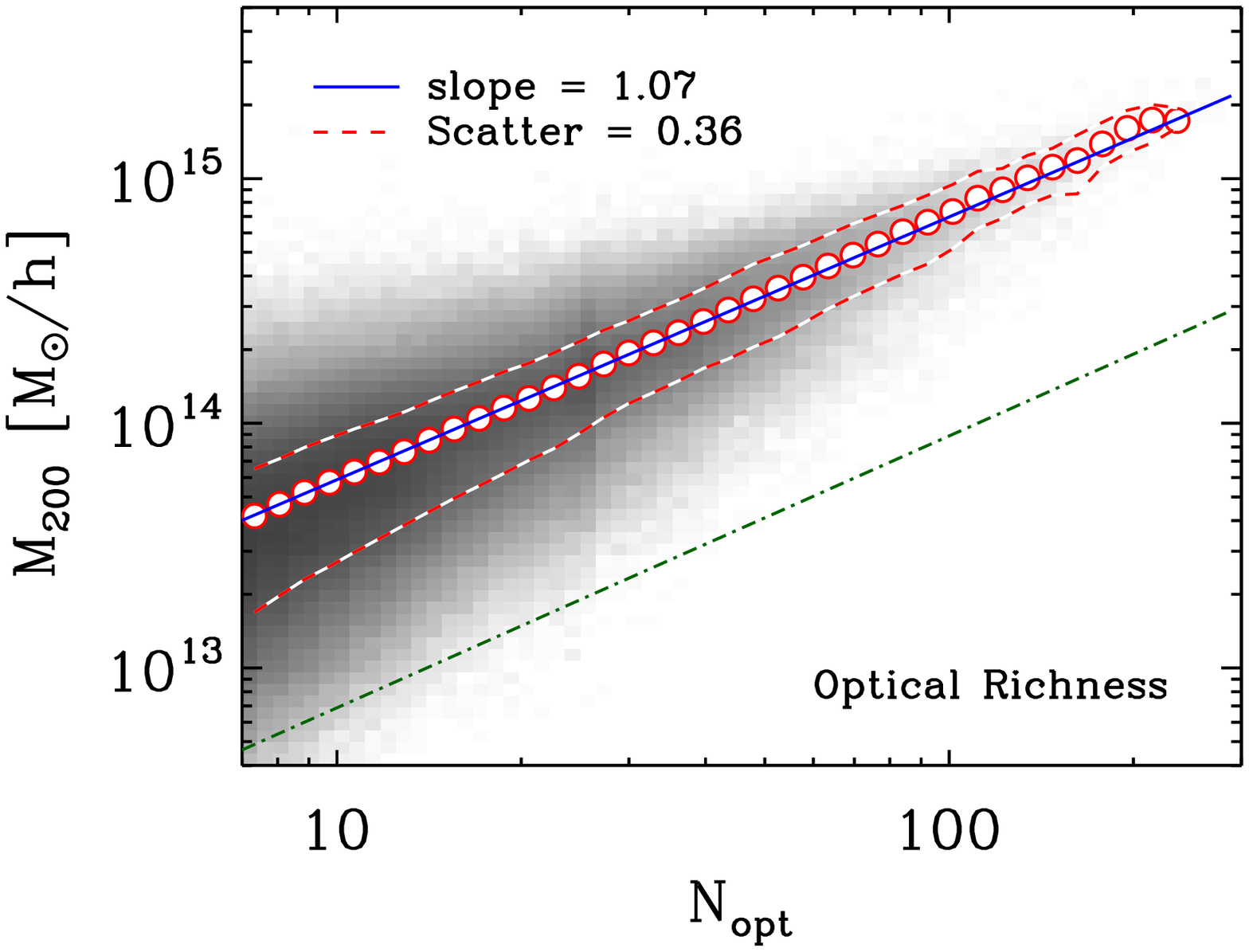}}
\resizebox{8.8cm}{!}{\includegraphics{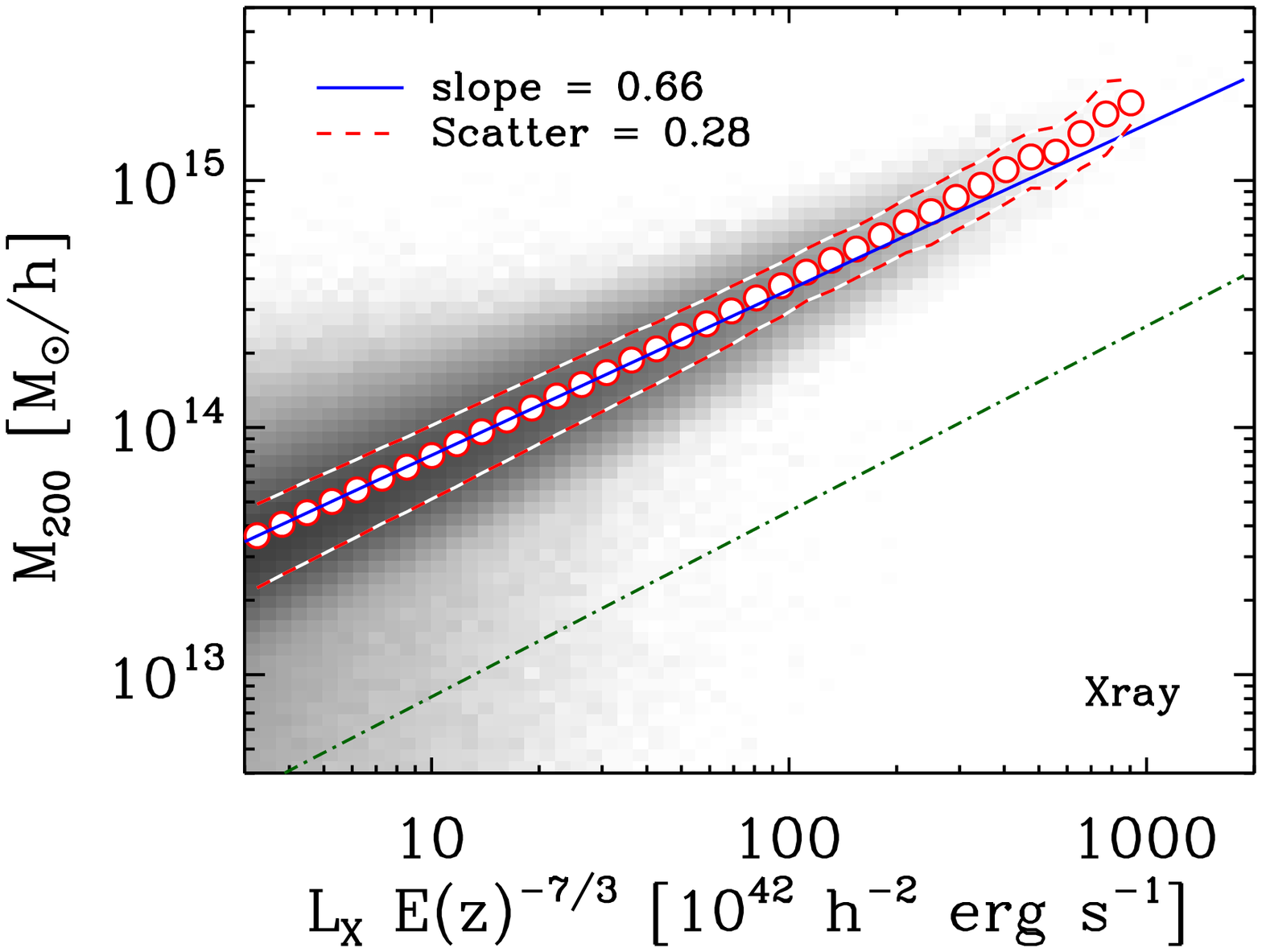}}\\
\resizebox{8.8cm}{!}{\includegraphics{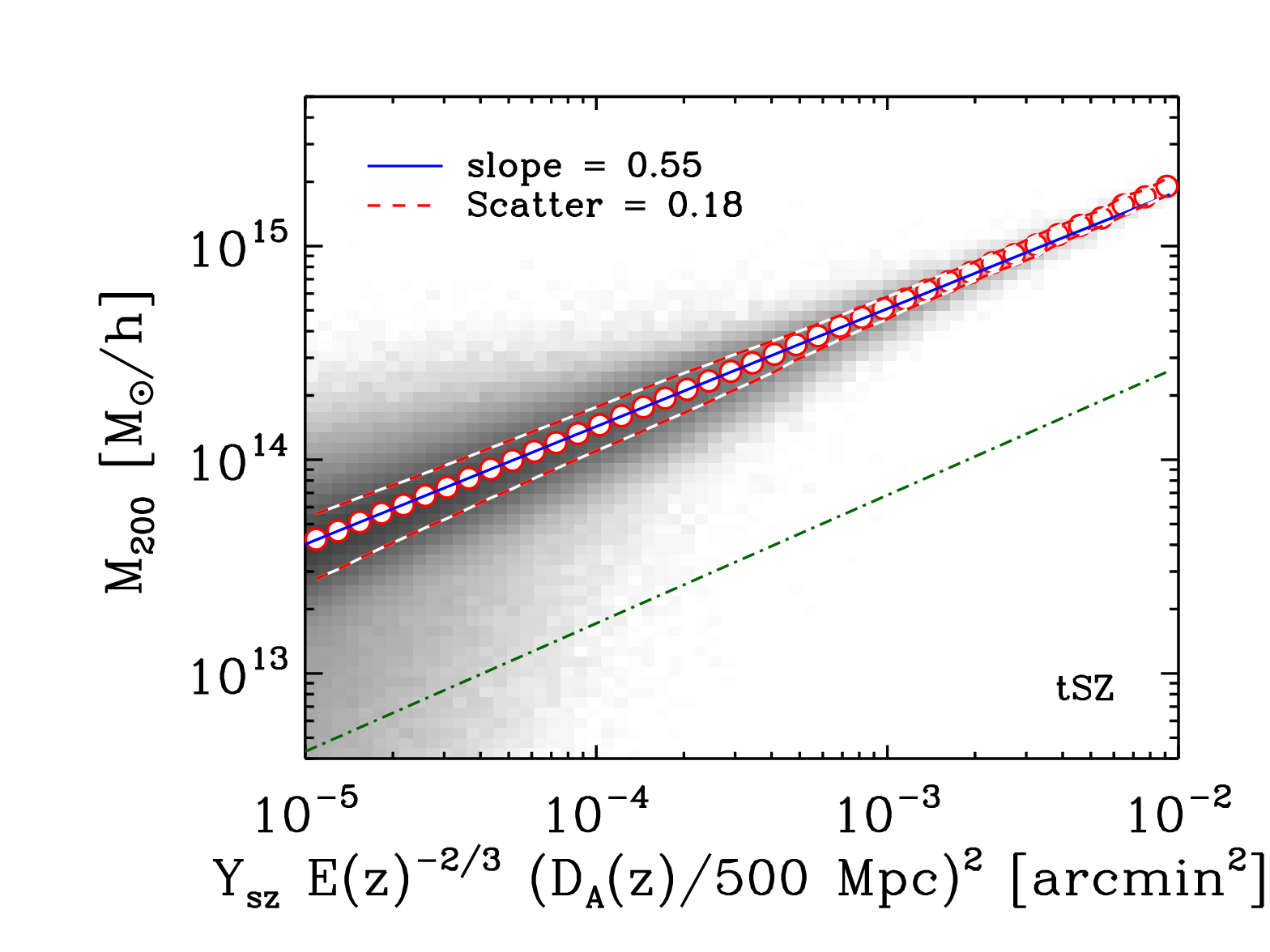}}
\resizebox{8.8cm}{!}{\includegraphics{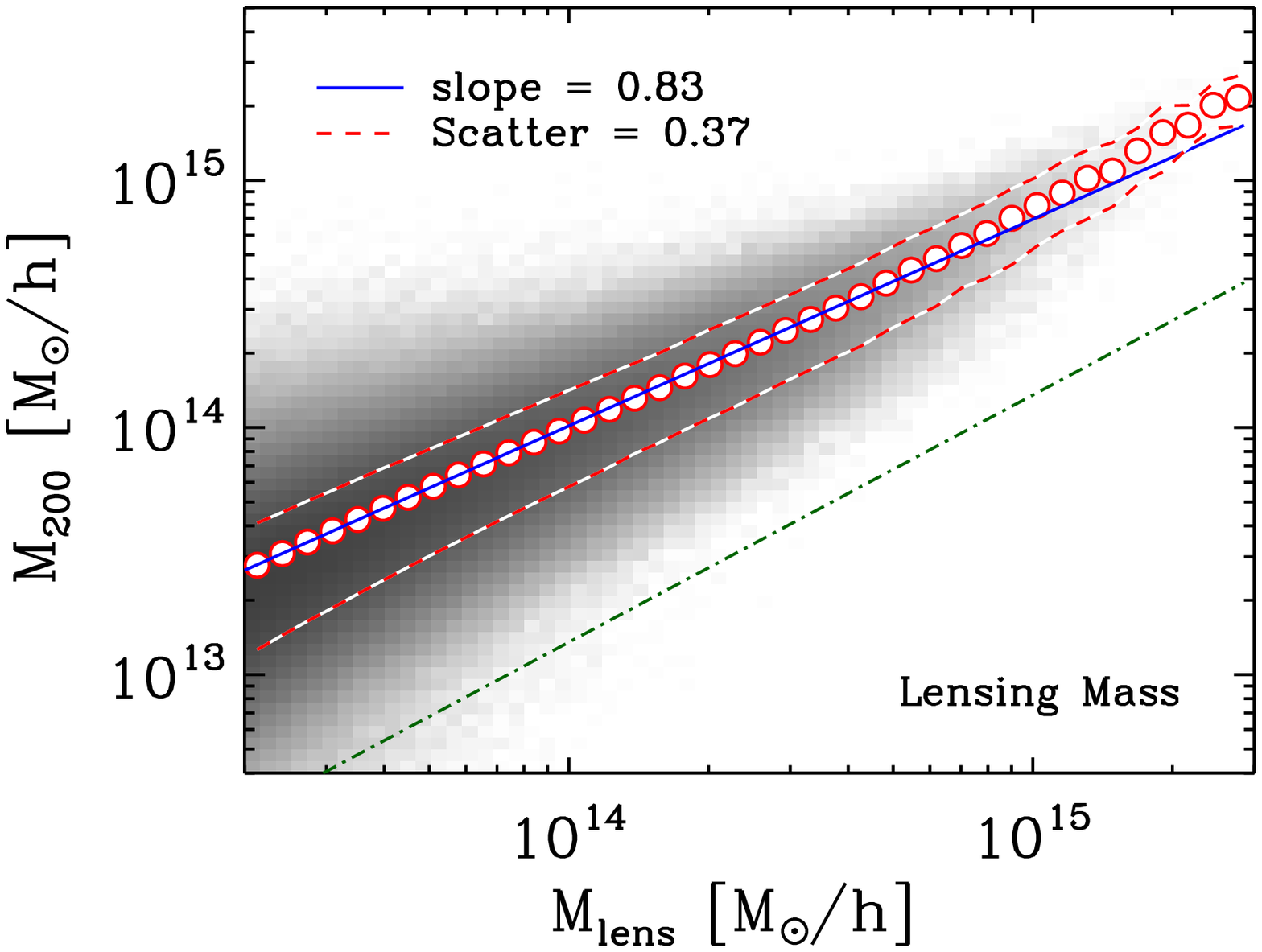}}
\caption{Virial masses ($M_{200}$) for a simulated volume-limited sample of
  optically detected galaxy clusters at $z=0.25$ as a function of our surrogate
  observables. Darker regions correspond to a larger number density of
  clusters.  Explicitly, \nopt$\,$ corresponds to a maxBCG-like optical
  richness, {\lx} to projected X-ray luminosity, {{\sz}} to projected tSZ
  signal and {\mlens} to weak lensing mass. All these signals are
  integrated values within the virial radius, calculated as described
  in the text, using cluster centres that are randomly displaced 30\%
  of the time from the true potential minima.  Dot-dashed lines
  indicate the expected `self-similar' scalings, while blue solid
  lines indicate the regressions (in log-log space) of halo mass
  against each surrogate. These have the
  slopes listed in each panel which differ from the
  self-similar expectation.  Circles show the mean virial mass for a series
  of narrow logarithmic bins of each surrogate, while dashed red lines
  indicate the region containing the central 68\% of the clusters in
  each bin. The fractional scatter in halo mass at given surrogate
  value is given in each panel.}
\label{plot:virialmass}
\end{figure*}

Our catalogue contains $594399$ objects with \nopt$\,$ above $10$,
corresponding roughly to haloes with $M_{200} >
4\times10^{13}\,h^{-1}{\rm M}_\odot$. There are $1988$ objects with
more than 100 members, corresponding to $M_{200} > 7\times10^{14}\,
h^{-1}{\rm M}_\odot$. For each cluster we compute associated surrogate
observables, the X-ray luminosity $L_{\rm X}$, the weak-lensing mass
$M_{\rm lens}$, and the tSZ flux $Y_{\rm SZ}$, by integrating the
corresponding 2D maps around the apparent (i.e. after `miscentering')
centre out to $\overline{R}(N_{opt})$. For each signal we subtract the
contribution of the background, which we estimate using an annulus of
radius $1.5\times \overline{R}(N_{opt}) < r < 2\times
\overline{R}(N_{opt})$.  Naturally, this is not the approach that one
would follow for individual well-observed clusters, where one can
directly identify the peak of the X-ray, tSZ or weak lensing signals
and estimate an individual virial radius from a profile built around
this centre. It is, however, closely analogous to the procedure
followed when estimating scaling relations (mean values of {\mlens},
{\lx} or {\sz} as a function of optical richness) by stacking large
samples of optically selected clusters.

In our analysis below, we consider two types of cluster sample that
mimic catalogues from large-scale optical and X-ray surveys. By
comparing results from these samples we hope to assess the impact of
the observational selection method on derived scaling relations
(c.f. section~3.2).  \cite{Koester2007a} show that, to a good approximation
and over a wide range of richness, their maxBCG catalogue can be
considered to be selected over a fixed volume of about 0.5~Gpc$^3$. We
thus take ``optically selected'' samples to be selected uniformly from
the full MXXL volume, independently of their properties.

Current large ``representative'' surveys of X-ray clusters are based
primarily on the Rosat All Sky Survey \cite[RASS, ][]{Voges1999} and on
 serendipitous discovery in fields observed for other reasons by the Rosat,
Chandra and XMM-Newton satellites. The largest compilations, such as the 1800
cluster {\MCXC} of \citet{Piffaretti2011}, are built by combining
subcatalogues each of which is effectively X-ray flux limited, and so
surveys a substantially larger volume for X-ray bright clusters than
for X-ray faint ones. For each subcatalogue, and so for the
compilation as a whole, the volume surveyed scales approximately as
$L_{\rm X}^{1.5}$ because the apparent luminosity decreases as distance
squared whereas the volume increases as distance cubed.  This
overrepresentation of luminous objects is known as Malmquist bias, and
we incorporate it in our ``X-ray flux limited'' samples by retaining all
objects in the MXXL volume but weighting each by the 3/2 power of {\lx}.

\subsection{Scalings with mass}

\begin{figure*}
\centering
\resizebox{17cm}{!}{\includegraphics{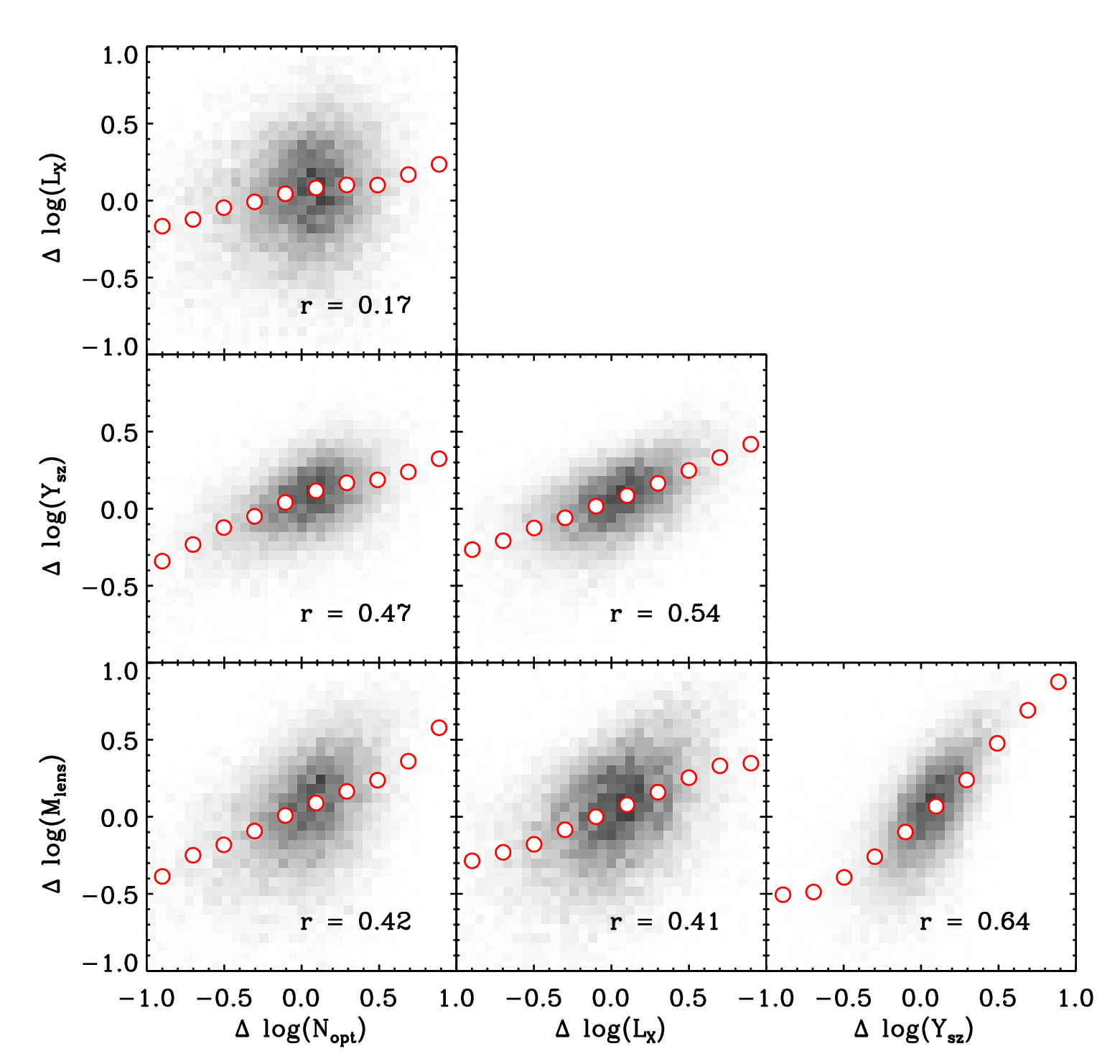}}
\caption{Correlations among deviations of observables in galaxy clusters with 
   mass in the range $4\times 10^{14}M_\odot < M_{200} < 1\times 10^{15}M_\odot$. 
   Data correspond to logarithmic deviations, i.e. 
   $\Delta\log(s) \equiv \log(s)-\langle \log(s) \rangle$, where the mean is 
   computed for clusters in narrow mass bins ($\Delta\log{M_{200}} = 0.2$). The
  intensity of the background 2D histogram is proportional to the number
  of haloes in the corresponding region of the plot, with a darker grey-scale
  indicating a larger number density of objects. Red circles correspond to
  the average $y$-value in bins along the $x$-axis. The linear correlation
  coefficient $r$ for each pair of observables is given in the legend
  of each panel.}
\label{correlations}
\end{figure*}

\begin{figure}
\centering
\resizebox{7.5cm}{!}{\includegraphics{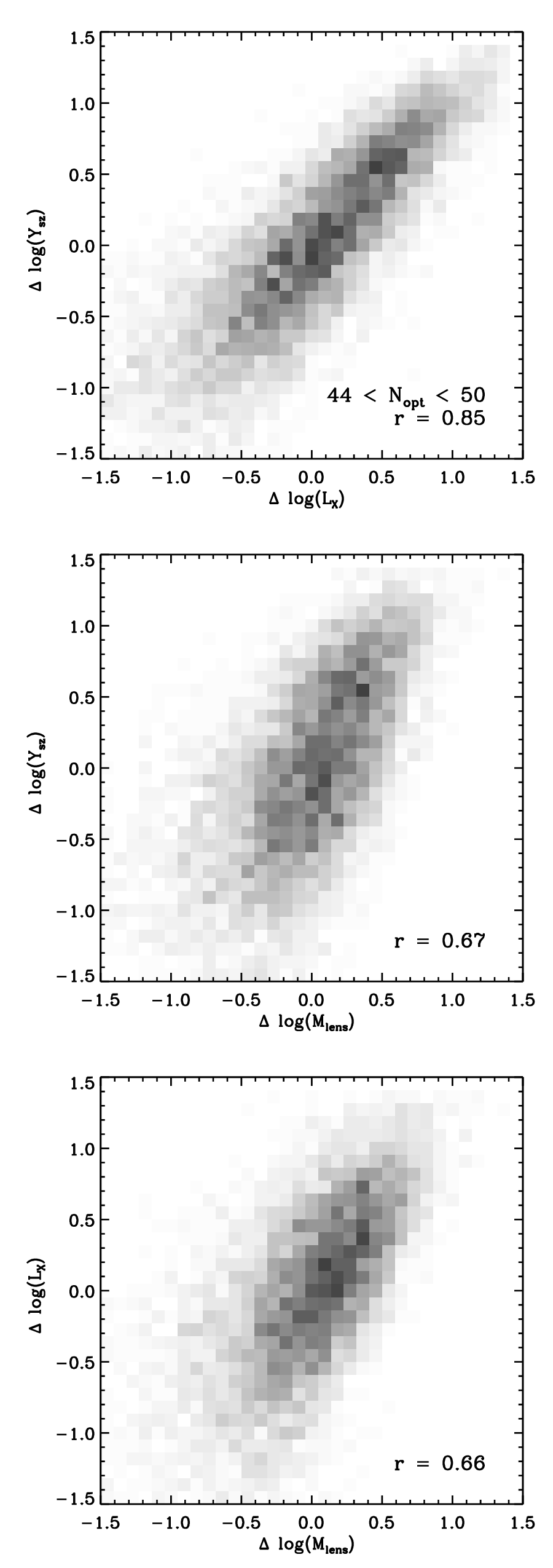}}
\caption{Correlations among deviations in {\lx}, {\sz} and {\mlens} for objects with
  optical richness in the range $44 < \nopt < 50$. The intensity of the
  background 2D histogram is proportional to the number of haloes in the
  corresponding region of the plot, with a darker grey-scale indicating
  a larger number density of objects. The linear correlation coefficient
  $r$ for each pair of observables is given in the legend of each panel.}
\label{correlations_nbin}
\end{figure}

In Fig.~\ref{plot:virialmass} we present the relations between
cluster virial mass $M_{200}$ and each of our four surrogate
observables for optically detected clusters. Specifically, {\nopt} 
corresponds to optical richness, {\sz}
and {\lx} to the projected tSZ and X-ray fluxes, and {\mlens} to weak
lensing mass, all integrated within the projected virial radius
corresponding (in the mean) to its estimated richness and surrounding
its apparent centre (i.e. after ``miscentering'' perturbations).  The
symbols in each panel correspond to the mean of the $M_{200}$
distribution at given value of the surrogate observable and the red
dashed lines contain the central 68\% of this distribution. The
straight blue lines show the linear fit to the individual cluster data
(in logarithmic space) which minimises the {\it rms} residuals in the
vertical direction.

For comparison, green dot-dashed lines indicate the scaling expected
naively given our assumptions about the relations between baryonic and
dark matter properties.  In the optical case, this comes directly from
the HOD model used to build the galaxy catalogues ($N_{\rm opt}\propto
M^{0.9}$), in the X-ray and tSZ cases from standard self-similar
scaling ($L_{\rm X}\propto M^{4/3}$ and $Y_{\rm SZ}\propto M^{5/3}$, respectively),
and in the case of lensing it is direct proportionality
($M_{\rm lens}\propto M$). In all four panels, the slope of the
measured regression, shown in the legend, is similar but not
identical to the expectation. These deviations can be explained
by the relative impact for the different surrogates of internal halo 
structure and of contamination along the line-of-sight, as well as of 
miscentering. All these can depend systematically on halo mass. 
The largest discrepancy is found 
for the lensing surrogate, followed by the optical richness. The smallest 
are found for the X-ray and tSZ signals. This is consistent with the fact
that the stronger the dependence of a surrogate on mass, the less sensitive 
it is to contamination and  to other projection effects, since these are
typically produced by less massive systems.

The scatter in halo mass at given value of an observable can be roughly
described by a log-normal distribution and depends weakly on the actual
value of the observable for those we study here. It is indicated in each
panel and ranges from $20$ to $40$\%. The tSZ signal shows the least scatter 
and the lensing the largest. The values we find are consistent with previous
studies, but note that our estimators are not optimal 
\citep[c.f.][]{Melin2006,Rykoff2012}. Note
also that our catalogues do not include all possible sources of scatter, so
even larger values may apply to real data. On the other hand, we consider that
our results should yield a reliable upper limit on the size of uncorrelated
projection effects, given the very large box size of the MXXL. In
particular, for tSZ our scatter estimates agree with those reported from full
hydrodynamical simulations of smaller volumes \citep{Kay2011,Battaglia2011};
and for optical richness, with those directly inferred from the data for 
optical clusters \citep{Rozo2009}. This is a reassuring confirmation that 
our assumptions are reasonable.

An interesting corollary of the considerable dispersion in these
relations is that it is unlikely that the cluster with the largest
value for any particular observable will actually correspond to the
most massive halo in the survey which identified it. We have
explicitly checked that this perhaps counterintuitive situation does
indeed hold.  In the insets of Fig.~\ref{mxxl}, we show the most
extreme cluster in our simulation as identified by each of our four
surrogate observables, i.e.~the cluster with the largest tSZ signal,
X-ray flux, gravitational lensing signal and optical richness
count. These clusters turn out all to be different. The cluster with
the largest richness is, in fact, the one with the largest $M_{200}$ at
$z=0.25$ and is notable also for the fact that it does not even have a
well defined centre. This makes clear that considerable care is needed
to draw cosmological inferences from the observed properties of the
most extreme cluster in any particular survey. Statistically
meaningful constraints can be obtained only with a complete and
accurate treatment of the scatter in the mass-observable relation,
including any possible dependence on cluster mass.

An important point for our subsequent analysis is that while some sources of
scatter affect primarily, or even exclusively, one specific observable, most
affect several simultaneously. For instance, at fixed $M_{200}$ the HOD is
expected to correlate with halo formation time because older and more relaxed
haloes tend to have more dominant BCGs. Since formation time correlates
strongly with concentration but only weakly with virial temperature, X-ray
luminosity is also expected to increase with halo age, whereas integrated
SZ-strength (and also lensing mass) should be age-independent. It has long
been clear observationally that at given richness more regular and relaxed
clusters (hence ``older'' clusters) do indeed have more dominant cD galaxies,
and it is now clear that they are also more X-ray luminous. In contrast,
variations in baryon fraction are expected to
affect {\lx} and {\sz} similarly, but to have little effect on {\mlens} and to
correlate in a model-dependent and uncertain way with \nopt. In addition,
orientation is expected to have little effect on the measured X-ray luminosity
of a cluster but to produce correlated variations in its measured SZ flux,
richness and lensing mass. Finally, misidentification of the centre and
misestimation of the virial radius of a given cluster will  induce variations in
all its observables. Generically, such effects imply that deviations
from the various mass-observable relations are not independent. Rather, there
is a non-zero covariance which reflects common sensitivities to halo
structure, orientation, environment and foreground/background superposition --
surrogates which are similarly sensitive to these factors are expected to
exhibit a high degree of correlation \citep[see also][]{Stanek2010}.

We quantify this effect in Fig.~\ref{correlations} which shows scatter
plots of the deviations from the mean at given $M_{200}$ in the
logarithms of the values of observables for individual clusters. Here
we include all clusters with $ 1\times 10^{15}M_\odot> M_{200} >
4\times 10^{14}M_\odot$. In each panel we give explicitly the Pearson
correlation coefficient, $r$, which characterises the correlation
between the deviations.

The strongest correlation is that between the deviations in lensing
mass and {\sz}, presumably because they are similarly sensitive to
cluster orientation, projection, miscentering and misestimation of
$R_{200}$.  The second strongest is between {\lx} and {\sz}, the two
quantities sensitive to our estimates of gas density and temperature. The
weakest is between richness and {\lx}, perhaps because the X-ray
luminosity is dominated by the dominant central concentration of
clusters while {\nopt} is influenced substantially by orientation and
projection effects.  The other three correlations are all of similar
strength.

While Fig.~\ref{correlations} illustrates the correlated scatter in
observables among clusters of given 'true' mass, the more relevant
correlations for the effects discussed in the next section are those
at fixed observed richness, {\nopt}. These are shown as scatter plots
for $44 < \nopt < 50$ in Fig.~\ref{correlations_nbin}. We have checked
and found quite similar results for other richness ranges.
The scatter in each observable is considerably larger at fixed {\nopt} 
than at fixed $M_{200}$ and the correlations are substantially stronger 
in Fig.~\ref{correlations_nbin} than in Fig.~\ref{correlations}, reaching 
$r=0.85$ for the particularly relevant case of {\sz} versus {\lx}.

Despite the different degrees of covariance among our surrogate observables,
we note that we measure a positive correlation in all cases. This means that a
cluster with an abnormally high signal in one surrogate is likely to have a
high signal in the other three as well, especially those where the relevant
correlation is strong. Hence, different observables do not provide independent
measurements of the true mass of any given cluster, and any analysis which
assumes that they do is likely to be at least partially in error.  Another implication is that a group of
clusters selected using of one of these observables will not form an
unbiased sample of the underlying cluster population with respect to any of
the other observables.  As a result, both the mean scaling relations for such
a sample and the scatter around these relations may differ from those for the
full underlying cluster population. We will see, for example, that the
mass-observable and observable-observable relations derived from an optically
selected cluster catalogue will differ in general from those derived from an
X-ray selected catalogue. 

\begin{figure}
\centering
\resizebox{8.5cm}{!}{\includegraphics{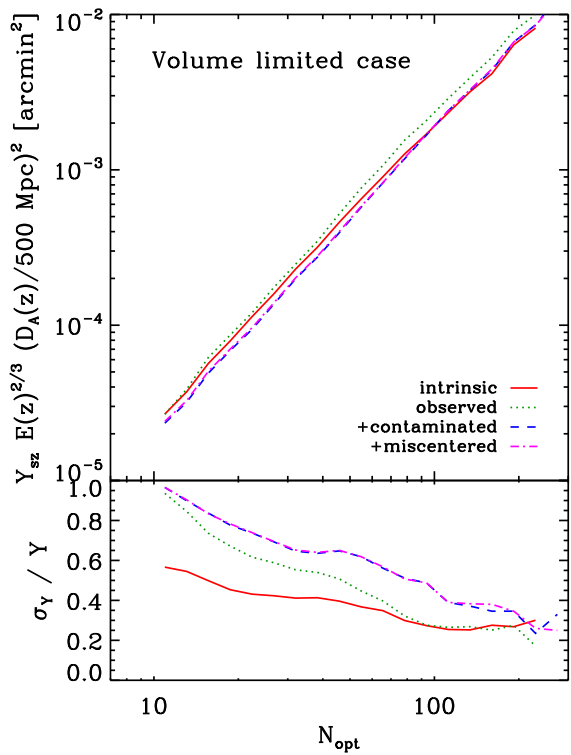}}
\caption{Thermal SZ signal {\sz} as a function of optical richness \nopt$\,$ for a
  volume-limited sample of $\sim 500,000$ clusters from the MXXL. The upper
  panel shows the mean relation and the lower panel the ratio of the scatter
  to the mean. In each panel the red curve shows the ``intrinsic relation''
  where both the tSZ signal and richness are computed in 3D within a sphere of
  radius $R_{200}$ centred on the potential minimum. For the ``observed''
  lines, both {\sz} and \nopt$\,$ are computed in projection about these true
  centres using the true $R_{200}$ and using only uncontaminated clusters, defined to be
  those where at least 90\% of the galaxies are members of a single FoF halo.
  The ``+contaminated'' lines refer to samples where this latter condition is
  eliminated and where the virial radius for each object is estimated
  iteratively from \nopt$\,$ as is done for real data. Finally, the
  ``+miscentred'' curves refer to samples where the projected position of the
  centre has been offset for 30\% of the clusters as described in the text.}
\label{plot:systematics1a}
\end{figure}

\subsection{Systematic effects in measured scaling relations}

\begin{figure}
\centering
\resizebox{8.5cm}{!}{\includegraphics{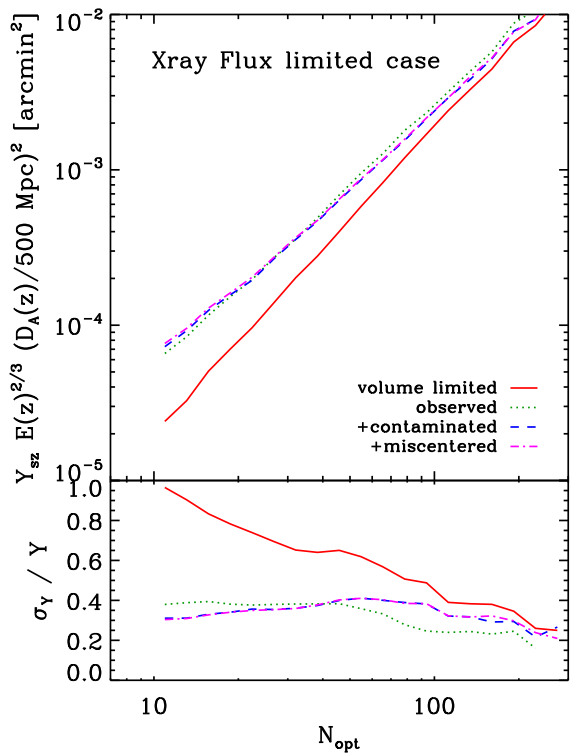}}
\caption{ SZ signal {\sz} as a function of optical richness \nopt$\,$ for a X-ray
  flux limited sample of $\sim 500,000$ clusters from the MXXL. This is
  directly analogous to Fig.~\ref{plot:systematics1a} and was constructed in
  the same way except each cluster is weighted by the 3/2 power of its X-ray
  luminosity. The ``observed'', ``+contaminated'' and ``+miscentered'' lines
  refer to the same cluster sets as in Fig.~\ref{plot:systematics1a},
  differing only because of this weighting. The red lines labelled ``volume
  limited'' repeat the ``+miscentered'' results from
  Fig.~\ref{plot:systematics1a}. Malmquist effects substantially enhance the
  amplitude of the mean relation, reduce the scatter, and make the relation
  insensitive to superposition and miscentering effects.}
\label{plot:systematics1b}
\end{figure}

We now consider how the thermal SZ signal is related to optical richness for
cluster samples selected in various ways. In Fig.~\ref{plot:systematics1a}
we display results where each MXXL cluster is assigned equal weight in order
to mimic volume-limited samples like the optically selected maxBCG survey.
For comparison, in Fig.~\ref{plot:systematics1b} we show results where each
cluster is weighted by $L_{\rm X}^{3/2}$, mimicking cluster samples like the {\MCXC}
which is constructed from a number of X-ray surveys, each of which is
effectively X-ray flux limited.  The upper panels in these plots show mean {\sz}
for clusters of given \nopt, while the lower panels show the {\it rms} scatter
about these relations expressed as a fraction of the mean signal. Lines in 
the upper panels thus represent the mean tSZ signals expected if clusters 
from optical or X-ray surveys are stacked as a function of their optical richness.

The solid red lines in Fig.~\ref{plot:systematics1a} show the
``intrinsic'' relation which is obtained if both {\sz}$\,$ and
\nopt$\,$ are calculated in 3D by integrating over a sphere centred on
the potential minimum and with radius $R_{200}$. The properties of
real clusters are, of course, measured from 2D maps. The dotted lines
labelled ``observed'' in both figures show the relations obtained for
uncontaminated clusters when both {\sz} and {\nopt} are integrated
over a disk around the potential minimum with radius given by the true
$R_{200}$. Here ``uncontaminated'' means that at least 90\% of the
galaxies contributing to {\nopt} have to lie in a single FoF
halo. Although, by definition, our estimates of {\nopt} and {\sz} both
increase for any individual object in going from 3D to 2D, they
increase by similar amounts, with the result that the apparent
relation does not change significantly over the full range of 
richness probed here. The scatter, on the
other hand, is greatly increased for poor clusters, where it can
double the intrinsic value, but is little affected for rich clusters.

For the dashed blue lines labelled ``+contaminated'' in
Figs.~\ref{plot:systematics1a} and~\ref{plot:systematics1b} we relax
the requirement that the clusters be uncontaminated. We also use an
observational estimate of $R_{200}$ for each cluster when estimating
its richness and its tSZ signal, as described earlier in this section.
Line-of-sight superposition effects then contribute to the values of
all the observables.  Because the superposed objects are usually of
lower mass (and hence lower temperature) than the main cluster, they
typically inflate {\nopt} (which scales approximately as $M^{0.9}$)
more than they do {\sz} (which scales approximately as $M^{5/3}$). As
a result, the ``+contaminated'' relations lie slightly to the right of
the ``uncontaminated'' ones. The effect is smaller in
Fig.~\ref{plot:systematics1b} than in Fig.~\ref{plot:systematics1a}.
The scatter is further increased by contamination in the volume-limited
case but is relatively little affected for flux-limited samples.

The final observational systematic we study is miscentering. The dot-dashed
purple curves labelled ``+miscentering'' in Figs.~\ref{plot:systematics1a}
and~\ref{plot:systematics1b} show the relations found when the centres
assigned to a random 30\% of the clusters are offset from their potential
minima as described above. This results in a surprisingly small shift in the
mean relation in both cases. Again, this is because such offsets induce shifts in the
estimated values of {\nopt} and {\sz} that are largely parallel to the mean
relation. 

These ``+miscentering'' curves give our most realistic estimate of the
relations expected for real clusters in the two cases. We repeat these
(dot-dashed purple) curves from Fig.~\ref{plot:systematics1a} as solid
red curves in Fig.~\ref{plot:systematics1b} in order to emphasise the
most important result of this section. The average {\sz} signal for
X-ray flux-limited samples is boosted by a factor of $3.5$ in the low
richness tail and by a factor of $1.25$ at high richness, relative to
the volume-limited case. This is a consequence of the strong
correlation between {\lx} and {\sz} at fixed {\nopt} which is visible
in the top panel of Fig.~\ref{correlations_nbin}.  In samples selected
above a limiting X-ray flux, clusters of given {\nopt} which are X-ray
underluminous are down-weighted and these tend also to be the objects
with the smallest {\sz} signals. This is a manifestation of Malmquist
bias.

\begin{figure}
\centering
\resizebox{8.5cm}{!}{\includegraphics{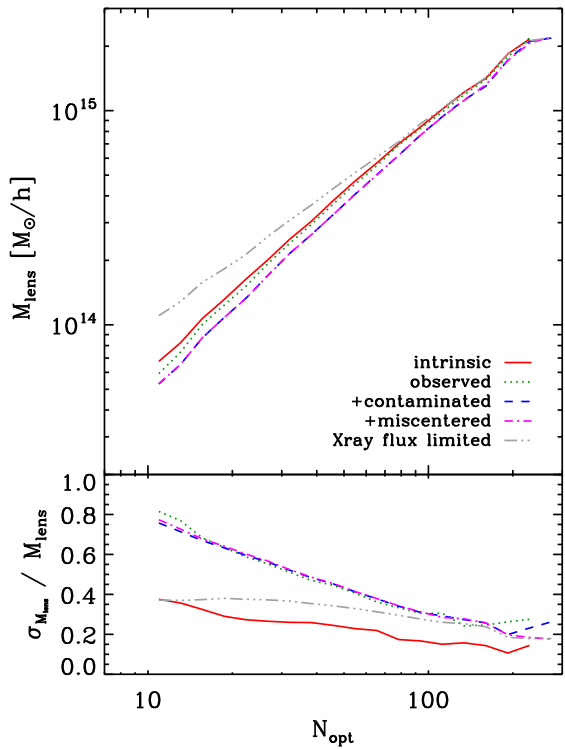}}
\caption{Mean lensing mass as a function of optical richness for clusters in
  the MXXL simulation. ``Lensing mass'' is defined as the mass within a
  properly centred sphere of radius $R_{200}$ ($M_{200}$, the red solid curve
  labelled ``intrinsic'') or projected within a circle of radius given by an
  \nopt-based estimate of $R_{200}$ and centred either on the true potential
  minimum (``observed'' and ``+contaminated'' cases) or on an offset centre in
  30\% of the cases (``+miscentred''). \nopt$\,$ is defined within these same
  regions. The upper panel shows the mean relations while the lower one gives
  the fractional scatter about these relations. The ``X-ray flux limited''
  curves refer to the ``+miscentred'' case with haloes weighted by $L_{\rm X}^{3/2}$. }
\label{plot:systematics2}
\end{figure}

A second important consequence of X-ray selection is that the scatter about the
mean {\sz}-{\nopt} relation is greatly reduced. The lower panel of
Fig.~\ref{plot:systematics1b} shows that we predict it to be only about half
that for a complete volume-limited sample of clusters. Although much of this
difference is due to a reduced sensitivity to contamination and miscentering,
the scatter is lower even than that shown in Fig.~\ref{plot:systematics1a} for
uncontaminated and properly centred clusters. If not corrected, this could
lead to over-optimistic estimates of the performance of mass estimators when
tested on X-ray selected cluster samples. Such samples are clearly more
homogeneous in internal structure at given mass than volume-limited samples.
We expect biases of this kind to be present in any cluster catalogue selected
according to a specific observable, and they must be corrected in order to
infer correctly, for example, the volume abundance of clusters as a function
of $M_{200}$, the quantity normally used to draw cosmological conclusions from
the cluster population.  Robust constraints require not only that the mean
transformation from observable to mass be determined accurately and without
bias, but also that the scatter between these quantities be known to high
precision.

Fig.~\ref{plot:systematics2} is analogous to the previous two figures
but now focuses on the relation between gravitational lensing mass and
richness. As was the case for {\sz}-{\nopt}, we find that the slopes
of the mean relations are similar in all the volume limited cases, and
that the effects of contamination and miscentering are quite modest.
In contrast to \cite{Johnston2007}, who find that miscentering
decreases the normalisation of the {\mlens}-{\nopt} relation by
15-40\% (see their Tables 3 and 8), it barely changes the mean
relation in our data. This is because of the strong correlation
between our estimators of optical richness and lensing mass. If the
center of a halo is misidentified, both {\nopt} and {\mlens} are
underestimated, and the change is, on average, almost parallel to the
mean relation. The stronger effect seen by \cite{Johnston2007} may
reflect their different estimator for the lensing mass, or possibly a
failure to account consistently for the implied change in richness.
We note that this richness reduction implies that at any given {\nopt},
fewer than 30\% of clusters will actually be miscentered, both because
cluster abundances increase steeply with decreasing richness, and
because poor clusters are more likely to be rejected by our algorithm
in favour of richer overlapping systems. Note also that the mean
lensing mass is biased high by the Malmquist effect in X-ray flux
limited cluster samples because at given richness more luminous
clusters are usually more massive.

The various biases examined here depend both on the intrinsic
properties of the cluster population and on the specific technique
used to measure each observable.  For instance, the smaller the
scatter between X-ray luminosity and optical richness, the smaller the
impact of Malmquist bias on the amplitude of the {\sz}-{\nopt}
relation. In the limiting case of no scatter the relations for
flux-limited and volume-limited samples would be identical. The same
would be true if there were no correlation between {\lx} and {\sz} at
fixed {\nopt}. The substantial effects found above are due to the
strong correlations we predict. The details of the
observational procedures matter because they can enhance or suppress
the impact of different aspects of the data. For instance, an optimal
filter based on expected cluster profiles will be more sensitive to
miscentering than a top-hat filter of large size; the shear profile
scheme which \cite{Johnston2007} used to estimate lensing masses
may be yet differently sensitive.  Such effects must be taken into
account properly if any particular survey is to place reliable
constraints on cosmological parameters. This also applies to
inferences from the properties of extreme objects. Cosmological
inferences require a full understanding of the scatter in the
observable-mass relation and an accurate knowledge of the selection
function, otherwise one may arrive at seriously erroneous conclusions.

\section{Scaling relations for Planck clusters}

\begin{figure*}
\centering
\resizebox{18cm}{!}{\includegraphics{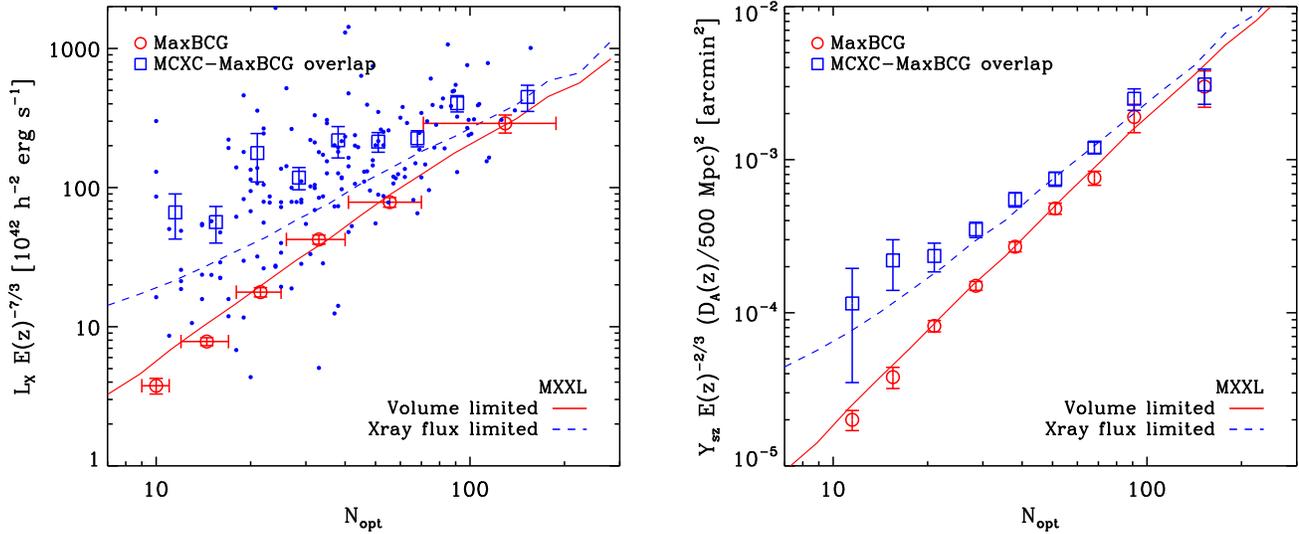}}
\caption{Comparison of scaling relations between observed and
  simulated clusters.  The left-hand panel shows X-ray luminosity as a
  function of optical richness. The red circles with error bars show
  the mean luminosity of stacks of MaxBCG clusters as measured by
  \citet{Rykoff2008b}, while blue points indicate individual maxBCG
  clusters which are also in the MCXC compilation of
  \citet{Piffaretti2011}. The right-hand panel shows the tSZ signal as
  a function of optical richness. Here, red circles indicate mean
  values for stacks of the entire maxBCG catalogue, while blue circles
  are for stacks of maxBCG clusters that are also in the {\MCXC}
  \citep{PlanckOptical2011}.  All observational data correspond to the
  equivalent of measurements within $R_{500}$ and at $z=0$.
  Predictions from the MXXL simulation are shown by red lines for a
  volume-limited sample and by blue lines for a X-ray flux-limited
  sample.  Note that the normalisations of our $L_{\rm X}$ and $Y_{\rm
    SZ}$ proxies are set by shifting the red lines vertically to give
  the best possible fit to the red circles in these two plots. }
\label{plot:planck1}
\end{figure*}

In the last section, we showed that cluster scaling relations are
strongly and systematically affected by the way in which cluster
samples are selected. This is a consequence of the substantial scatter
in mass-observable and observable-observable relations, and the fact
that deviations of different observables from the mean relations are
strongly correlated because of common sensitivities to cluster
structure, orientation, environment, and line-of-sight projection. The
resulting distortion of scaling relations depends on how clusters are
detected and their observables measured, so precise correction
requires detailed modelling of each individual experiment. Our
catalogues based on surrogate observables are nevertheless realistic
enough to examine whether effects of this kind might explain some
apparent inconsistencies recently highlighted by the {\planck}
Collaboration.

There are three pieces to the puzzle presented by the {\planck}
collaboration.  First, the mean tSZ signal for stacks of maxBCG
clusters of given optical richness is about half of that predicted by
scaling relations derived from the much smaller {\REXCESS} sample for
which individual X-ray profiles are available \citep{Bohringer2007}.
The hot gas structure of the {\REXCESS} sample is well described by
the almost self-similar scaling of a ``universal'' pressure profile
which resembles that predicted by hydrodynamical simulations of
cluster formation \citep{Arnaud2010}.  The {\REXCESS} scaling
relations do, however, agree well with the mean {\sz} measured when
clusters from the large {\MCXC} compilation \citep{Piffaretti2011} are
stacked as a function of {\lx} \citep{PlanckXray2011}. Note that
comparison with the stacked {\sz}-{\nopt} relation for maxBCG sample
requires an estimate of cluster mass as a function of {\nopt}, for
which \citet{PlanckOptical2011} used weak lensing results for stacked
maxBCG clusters \citep{Johnston2007, Rozo2009}. 

The second piece of the puzzle is that if the maxBCG sample is
restricted to clusters which are also in the {\MCXC}, the stacked
{\sz}-{\nopt} relation lies well above (a factor of 2 at ${\nopt}=50$)
that for the sample as a whole and appears consistent with the
{\REXCESS} scaling relations. The third and final piece is that the
relation between mean {\sz} and mean {\lx} signals for stacks of the
full maxBCG binned by {\nopt} is consistent both with that between
mean (stacked) {\sz} and {\lx} in the {\MCXC} and with that between
{\sz} and {\lx} for individual clusters in the {\REXCESS} sample. All
these facts led the {\planck} Collaboration to speculate that a subset
of optically detected clusters might have very weak tSZ and X-ray
signals, presumably because they contain many galaxies but rather
little hot gas.

We now examine these issues using suitably selected cluster catalogues from
the MXXL. To mimic the full maxBCG catalogue, we use volume-limited samples
\citep[c.f.][]{Koester2007a} and we measure a ``maxBCG-like'' richness for
each as detailed in Section 3.1. To mimic the {\MCXC}, we will use ``X-ray flux
limited'' samples which weight each object by $L_{\rm X}^{3/2}$, since
\citet{Piffaretti2011} built the {\MCXC} by combining a number of X-ray surveys,
most of which are effectively flux-limited. We are also interested in
the overlap between these two observational samples. At high X-ray luminosity
clusters are detected in the RASS to distances beyond the limit of the maxBCG
catalogue, so that the combined sample is effectively volume-limited. At low
X-ray luminosity, on the other hand, the maxBCG limit is well beyond the
distance at which RASS can detect clusters and the overlap sample is
flux-limited. As we will see below, the effects of this change in sample
selection are directly visible in the behaviour of the stacked {\sz} signal of
the overlap sample.

We begin by considering the mean X-ray luminosity and mean tSZ signal as
functions of optical richness for MaxBCG clusters. These data are displayed in
Fig.~\ref{plot:planck1} as red circles. In the left panel, we show X-ray
measurements from \cite{Rykoff2008b} which correspond to the average X-ray
luminosity within $R_{500}$ for stacks of RASS maps centered on MaxBCG
clusters. We convert their luminosities, originally measured within $R_{200}$,
to ones measured within $R_{500}$ by multiplying them by $0.91$. We also scale
these values to their $z=0$ equivalent by assuming a self similar scaling of the 
luminosities. In the right panel we show the average tSZ signal in {\planck} 
maps within 
$R_{500}$ for the same richness-binned MaxBCG clusters. Note that the 
mass-richness relation from \cite{Rozo2009} was used to set the size of
the matched filter when measuring the {\sz} signals, but that this has little
impact on the result and is unimportant for our discussion here.

We now compare these results to our simulation. Red lines show the mean
relation predicted by ``observational'' catalogues which include both
contamination and miscentering. They have been shifted vertically to fit the
maxBCG data as well as possible, thus determining the scaling coefficients
used consistently in our {\lx} and {\sz} proxies throughout this paper.  After
this normalisation, the $Y_{\rm SZ}-N_{200}$ and $L_{\rm X}-N_{200}$ relations
are predicted remarkably well. The latter relation in the MXXL is slightly 
shallower than observed. This reflects the well known fact that
the observed {\lx}-$T_{\rm X}$ relation is steeper than that expected from the
self-similar model on which we built our surrogates, apparently indicating a
systematic variation in gas fraction or concentration with cluster mass.  The
agreement nevertheless appears good enough to validate our approximate
modelling of gas physics, suggesting that it is adequate to capture the
statistical correlations underlying the influence of selection bias on
cluster scaling relations.

The second set of observations in Fig.~\ref{plot:planck1} refer to the set
of $189$ maxBCG clusters which are also in the {\MCXC}. Blue dots in the left-hand
panel show {\lx} as a function of {\nopt} for the individual clusters, whereas
blue squares in the both panels give average values for stacks of
these clusters around their maxBCG centres, using the same {\nopt} bins as for
the full maxBCG sample \citep{PlanckOptical2011}. In both cases we plot the
$z=0$ equivalent value, using the measured redshift for each cluster and assuming 
a self-similar scaling of the signals. This allows a consistent comparison of all
datasets. 

As noted by the {\planck} Collaboration, stacked tSZ fluxes are
systematically larger for {\MCXC} clusters than for the full MaxBCG
sample, except possibly for the richest systems. The left panel
ofFig.~\ref{plot:planck1} indicates that their X-ray luminosities are
also systematically higher, again with the possible exception of the
richest clusters.  Blue dashed lines in both panels indicate the mean
relations we predict for X-ray flux limited samples. The Malmquist
bias offset from the volume-limited relation explains part of the
difference between the blue squares and the red circles in the left
panel, and it explains almost completely the discrepancy in the right
panel. As discussed in the last section, the latter is caused by a
strong correlation between the deviations of individual clusters from
the mean {\lx}-{\nopt} and {\sz}-{\nopt} relations.  This causes
Malmquist bias to propagate from X-ray selection into the {\sz}-
{\nopt} relation\footnote{\cite{Rykoff2008} give extensive discussion
  of various bias effects when the maxBCG catalogue is combined with
  X-ray cluster surveys.}. This suggests that the correlated scatter
between {\lx} and {\sz} at given {\nopt} is well represented in our
model, but also that there are sources of scatter which only affect
{\lx} that are not accounted for in our analysis which would increase
the difference between X-ray scaling relations derived from volume and
X-ray flux limited samples without affecting the {\sz} signal.

Finally, we reiterate that both relations have considerable
scatter. At given {\nopt} our model indicates that the fractional
uncertainty in {\lx} and {\sz} for volume-limited samples is about
$40\%$ for {\nopt} $\sim 200$ and rises to $130\%$ for {\nopt} $\sim
10$.  These numbers are broadly consistent with the intrinsic scatter
reported by \citet{PlanckOptical2011} for the SZ measurement of MaxBCG
clusters. The corresponding fractional scatters are about $40\%$ for
both {\lx} and {\sz} in our X-ray flux limited samples.

\begin{figure}
\centering
\resizebox{8.5cm}{!}{\includegraphics{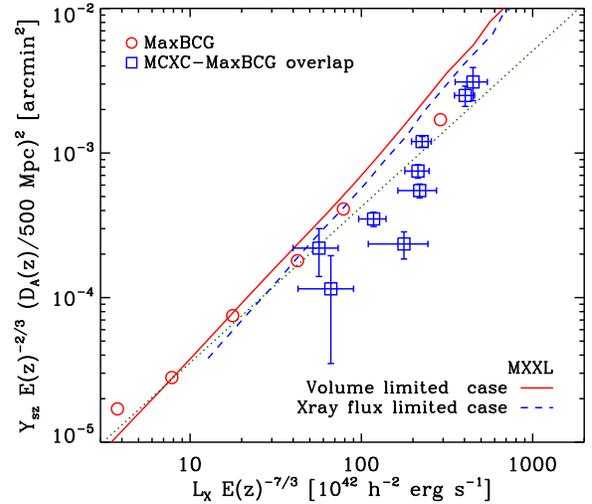}}
\caption{Average tSZ flux as a function of average X-ray luminosity for clusters
  stacked according to optical richness. Red solid and blue dashed lines give
  MXXL results for our volume-limited and X-ray flux limited samples,
  respectively and refer to the same set of {\nopt} bins. Both {\lx} and {\sz} 
  are  larger for the flux-limited sample, but the shift is almost parallel to
  the  mean relation so that Malmquist bias has little visible effect. Red
  circles are taken from \citet{PlanckOptical2011} and refer to maxBCG clusters,
  whilst blue squares are the results for the subset of clusters present in the 
  {\MCXC} catalogue. In both cases the data are binned according to {\nopt}. 
  Error bars indicate bootstrap uncertainties in the mean {\lx} and {\sz} signals
  in each richness bin. The green dotted line shows the predictions of the 
  X-ray model build by \citet{PlanckOptical2011}.}
\label{plot:planck2}
\end{figure}

We now move to another observable scaling relation reported by {\small
PLANCK}: that between {\sz} and {\lx}. We display simulation results for both
volume-limited and X-ray flux limited catalogues in Fig.~\ref{plot:planck2}.
Clusters were stacked as a function of {\nopt} and then the mean {\sz} of each
stack was plotted against its mean {\lx}. Although both {\sz} and {\lx} are
substantially larger at given {\nopt} in flux-limited stacks, the two relations
are almost identical. Malmquist bias has a negligible impact because the 
shifts are almost parallel to the mean relation.

The same relation can be constructed for the (almost) volume-limited
maxBCG catalogue by stacking RASS and {\planck} data for clusters
binned by \nopt. This exercise was carried out by
\citet{PlanckOptical2011} and their result is overplotted in
Fig.~\ref{plot:planck2}. We also include, as a green dotted line, the
relation predicted by the X-ray model built by
\citet{PlanckOptical2011} based on observations of {\REXCESS} clusters
and an observational {\nopt}-{\mlens} calibration.  Despite the
relations being built from different samples, they are very similar to
each other and to our MXXL predictions, showing not only that the
{\sz}-{\lx} relation is insensitive to details of sample definition,
but also that our proxies continue to represent the observables well
in this plane.

An {\sz}-{\lx} relation for the effectively flux-limited
{\MCXC}-maxBCG overlap is also shown in Fig.~\ref{plot:planck2}. For
the richest optical bins, it is compatible with all other relations,
but the {\lx} signal is considerably larger for poor clusters.  The
relatively small number of systems per bin leaves room for this to be
a statistical fluctuation (and indeed, we find that the median per bin
agrees much better with the expected relation). However, if real, it indicates that the differences in {\sz} and {\lx}
at fixed {\nopt} cannot be explained by the existence of a subpopulation
devoid of hot gas, as suggested by \citet{PlanckOptical2011}, since this would
move the points in Fig.~\ref{plot:planck2} along a line of slope unity and so could
not bring them back onto the relation for the full maxBCG sample. On the other hand, all data can
be simultaneously explained by Malmquist bias, together with an extra
source of scatter affecting only {\lx} (e.g. the presence or absence
of cool cores in the gas distributions). Note that this requires correlated
scatter in observables of the kind illustrated in Fig.~\ref{correlations} and 
Fig.~\ref{correlations_nbin} and does not work in simpler treatments of the 
same puzzle, as that of \cite{Biesiadzinski2012} which does not fully incorporate 
all such correlations.

Given that our simulation appears to reproduce consistently all the
measured observable-observable relations for the maxBCG and {\MCXC}
catalogues, it is interesting to discuss {\it why} the {\planck}
collaboration's modelling gave a prediction which was inconsistent
with the measured {\sz}-{\nopt} relation for the maxBCG sample.  The
answer appears to lie in the way cluster masses were used to relate
the population identified in X-ray surveys to the maxBCG
clusters. Since values of {\nopt} are not available for most of the
clusters in bright and well studied X-ray samples, this richness
measure must be calibrated against another observable if the hot gas
properties of such samples are to be used to predict the mean {\sz}
signal from maxBCG stacks. X-ray luminosity could have been used,
since \citet{Rykoff2008} provide mean {\lx} values as a function of
{\nopt}, and the {\sz}-{\lx} relation is not only theoretically robust
but also observationally well determined (see
Fig.~\ref{plot:planck2}). This route
($\nopt\rightarrow{\lx}\rightarrow{\sz}$) predicts mean {\sz} values
for maxBCG stacks which agree well with those measured
directly. However, the {\planck} collaboration decided instead to
follow a different route ($\nopt\rightarrow M_{500}\rightarrow{\sz}$)
using the mean weak lensing masses measured as a function of {\nopt}
for stacked maxBCG clusters by \citet{Johnston2007} and \citet{Rozo2009}
who in addition corrected {\mlens} upwards to account for an
improved model for the redshift distribution of the source galaxy population.
Both these studies made substantial upward corrections to the directly
measured  mean masses to account for the effects of 
line-of-sight contamination and miscentering, but they failed to make
consistent corrections to {\nopt}, thus ignoring the correlated
deviations of individual clusters from the mean {\nopt}-$M_{200}$ and
{\mlens}-$M_{200}$ relations which are visible in the lower left panel
of Fig.\ref{correlations}. Thus, the reported scaling 
relations correspond to the relationship between the true mass of
a cluster and its miscenterd optical richness, and so predict the {\sz} signal
expected for well-centered clusters as a function of 
their miscentered richness. These predictions should not be compared
against Planck data, unless the {\sz} is corrected for miscentering
in a similar way to the weak lensing data.

An alternative is to use the uncorrected measurements, because
the correlation among observables results in an apparent {\nopt}-{\mlens}
relation almost identical to that of well-centered and uncontaminated
clusters. In fact, if the original ``raw'' masses obtained by
\citet{Johnston2007} are used instead of the ``corrected'' masses to
predict mean {\sz} as a function of {\nopt}, the discrepancy with the
directly measured values almost disappears.

\section{Conclusions}

Throughout this paper, and in particular in the previous section, we
have emphasised the importance of understanding how survey methods
influence the scaling relations measured in the galaxy cluster
catalogues they produce. We have argued that this is crucial both for
proper statistical analysis of the physical properties of the cluster
population and for deriving meaningful cosmological constraints from
the estimated masses of the extreme clusters identified in any given
survey.

In order to illustrate these points, we have used the dark matter
distribution in the MXXL simulation, the largest high-resolution
cosmological calculation to date, to construct sky maps from which
clusters can be catalogued using proxies for four different
observables: optical richness as measured in deep photometric redshift
surveys, X-ray luminosity, thermal Sunyaev-Zeldovich (tSZ) signal, and
weak lensing mass.  Although our treatment of these observables is
necessarily simplified, it is sufficient to explore the scatter in the
relation between each observable and cluster mass, as well as the
correlations between the scatter for different observables caused by
common sensitivities to the internal structure, orientation,
environment and background contamination of clusters. This is
essential to understand the systematic biases imposed by specific
observational strategies for detecting clusters and measuring their
properties.

We employed these catalogues to show that there are a number of
effects that systematically alter the slope, amplitude and scatter of
scaling relations among the observables. Structural complexities,
orientation variations, superposition both of surrounding large-scale
structure and of unrelated foreground/background objects, and
miscentering all increase the scatter in the {\sz}, {\lx}, {\mlens}
and {\nopt} signals for given cluster mass. Relations between these
observables, for example, the {\mlens}-{\nopt} or {\sz}-{\lx}
relations, can, however, be much less affected because clusters
scatter roughly parallel to the mean relation. In addition, Malmquist
effects in flux-limited surveys not only bias the amplitude and reduce
the scatter in the mass-observable relation for the observable used
to select the sample, but also in those for other observables which
have correlated scatter.  The strength of such effects depends
substantially on survey strategy and on the operational definition of
the observables.

Ignoring these bias effects can lead to serious difficulties in
interpreting cluster data. As an example, we have considered a
discrepancy recently highlighted by the Planck collaboration which 
concerns the mean tSZ and X-ray signals
measured for stacks of clusters identified from optical and X-ray
surveys. Both signals are lower for optically selected clusters than
predicted for their weak lensing estimated masses by a model which
fits both individually observed and stacked X-ray-selected
clusters. Our results suggest that the data are nevertheless in good
agreement with predictions for a concordance $\Lambda$CDM universe,
even if the gas properties of clusters are assumed to scale in a
simple self-similar way with cluster mass. The discrepancy appears to
reflect Malmquist bias propagating from the X-ray luminosities to the
tSZ signal through covariance in their scatter at fixed cluster mass.
Malmquist bias has rather little impact on the mean $Y_{\rm
SZ}-L_{\rm X}$ relation, since clusters scatter almost along it. The
discrepancy appears to have been exacerbated by applying a substantial
miscentering correction to the mean {\mlens} for the stacked clusters
without applying a corresponding correction to the mean values of the
other relevant observables. Our model suggests that together these
effects may resolve the apparent puzzle.

Although our analysis appears to explain the discrepancy both
qualitatively and quantitatively, our explanation should still be regarded as
provisional because of the detailed dependence of the effects on how the
observables were obtained from the observational data. A firmer
conclusion can only be reached through considerably more detailed
modelling of the particular surveys involved. This should address not
only survey design and cluster identification issues, but also the
specific algorithms (matched filters, etc.) used to measure the
observables.  Additional uncertainty comes from our schematic
treatment of the baryonic physics, which undoubtedly misses important
aspects of the relation between the visible material and the
underlying mass. It is nevertheless clear that precision cosmology
with clusters will be impossible without carefully tailored surveys
with calibration strategies that fully account for the
multidimensional scatter between all the relevant observables and the
fiducial cluster mass. Furthermore, linking the observations to the
underlying cosmological model will require simulations that model all
these statistical and astrophysical aspects to the required level of
precision.  Even with its limited treatment of the relevant
astrophysics, the remarkable size and statistical power of the MXXL
gives a foretaste of what should be possible in the future.

\section*{Acknowledgements}

We thank the staff at the J\"ulich Supercomputer Centre in Germany for
their technical assistance which helped us to successfully complete
the MXXL simulation. We would like to thank the referee, James
Bartlett, for a insightful report. We also acknowledge useful
discussions with E. Rozo and C. Hernandez-Monteagudo. The initial 
conditions software was developed and tested on COSMA-4 which is part of 
the DiRAC Facility jointly 
funded by STFC, the Large Facilities Capital Fund of BIS, and
Durham University. RA and SW are supported by Advanced Grant 246797 
``GALFORMOD'' from the European Research Council. VS and SW acknowledge 
support by the DFG Collaborative Research Network TR33 ``The Dark Universe''.

%----------------------------------------------
\bibliographystyle{mn2e} \bibliography{ext}
%---------------------------------------------------------------------

\label{lastpage} \end{document}